%% file: longrmholes.tex
\documentclass[11pt]{article}
\pdfoutput=1
\usepackage{jheppub}
\usepackage{tabu}
\usepackage[usenames,dvipsnames,table]{xcolor}
\usepackage{graphicx,amsmath,amssymb,amsthm,multirow,array}
\usepackage[bbgreekl]{mathbbol}
\usepackage[numbers,sort&compress]{natbib}
\usepackage{verbatim}
 \usepackage{color} 
\usepackage{ccaption}
  \captionnamefont{\bfseries}
  \captiontitlefont{\small\sffamily}
  \captiondelim{: }
  \hangcaption
  \renewcommand{\figurename}{Fig.}

\usepackage{verbatim}
\usepackage{graphics, color}
\usepackage{graphicx}
\usepackage{amsmath}
\usepackage{amssymb}
\usepackage{amsthm}
\usepackage{amsfonts}
\usepackage[usenames,dvipsnames]{xcolor}
\usepackage[normalem]{ulem}

\DeclareFontEncoding{U}{}{}
\DeclareFontFamily{U}{bbold}{}
\DeclareFontShape{U}{bbold}{m}{n}
 {  <5> <6> <7> <8> <9> gen * bbold
   <10> <10.95> bbold10
  <12> <14.4> bbold12
 <17.28> <20.74> <24.88> bbold1 7
  }{}
\DeclareSymbolFont{bbold}{U}{bbold}{m}{n}
\DeclareSymbolFontAlphabet{\mathbbold}{bbold}
\DeclareMathAlphabet{\mathbbmsl}{U}{bbm}{m}{sl}
\renewcommand{\thefigure}{\arabic{figure}}

\def\nref#1{(\ref{#1})}
\newcommand{\told}{\mathbbmsl{t}}
\newcommand{\ext}{\emph{extremal} }
\newcommand{\Hold}{\mathbbmsl{H}}
\newcommand{\sold}{\mathbbmsl{s}}

\definecolor{rust}{rgb}{0.8,0.2,0.2}
\definecolor{purple}{rgb}{0.8,0.1,0.9}
\definecolor{olivegreen}{rgb}{0,0.52,0.17}

\title{%
  Inside Out: Meet The Operators Inside The Horizon \\
  \large On bulk reconstruction behind causal horizons \\
  }
\author[a]{Ahmed Almheiri}
\author[b]{\!\!, Tarek Anous}
\author[a]{\!\!, Aitor Lewkowycz}
\affiliation[\,a]{Stanford Institute for Theoretical Physics, Department of Physics, \\
Stanford University, Stanford, CA 94305, USA}
\affiliation[\,b]{ Department of Physics and Astronomy, University of British Columbia, 6224 Agricultural Road, Vancouver, B.C. V6T 1Z1, Canada
}

\emailAdd{almheiri@stanford.edu}   \emailAdd{tarek@phas.ubc.ca} \emailAdd{lewkow@stanford.edu}
\abstract{
Based on the work of Heemskerk, Marolf, Polchinski and Sully (HMPS), we study the reconstruction of operators behind causal horizons in time dependent geometries obtained by acting with shockwaves on pure states or thermal states. These geometries admit a natural basis of gauge invariant operators, namely those geodesically dressed to the boundary along geodesics which emanate from the bifurcate horizon at constant Rindler time.  We outline a procedure for obtaining operators behind the causal horizon but inside the entanglement wedge by exploiting the equality between bulk and boundary time evolution, as well as the freedom to consider the operators evolved by distinct Hamiltonians. This requires we carefully keep track of how the operators are gravitationally dressed and that we address issues regarding background dependence. We compare this procedure to reconstruction using modular flow, and illustrate some formal points in simple cases such as AdS$_2$ and AdS$_3$.

}

\begin{document}
\maketitle
\input all.sections

\acknowledgments
We would like to thank William Donnelly, Monica Guica, Veronika Hubeny, Daniel Jafferis, Don Marolf, Eric Mintun, Mukund Rangamani, and Aron Wall  for useful discussions. T.A. is supported by the Natural Sciences and Engineering Research Council of Canada, and by grant 376206 from the Simons Foundation.  A.L. acknowledges support from the Simons Foundation through the It from Qubit collaboration. A.L. would also like to thank the Department of Physics and Astronomy at the University of Pennsylvania for hospitality during the development of this work.



\newpage
\bibliographystyle{jhep}
\bibliography{biblio}

\end{document}

%% file: all.sections.tex
\section{Introduction}
\label{sec:intro}

In AdS/CFT, bulk locality is an emergent property of the boundary CFT. While locality can be probed directly from the correlators of boundary local operators \cite{Heemskerk:2009pn,Maldacena:2015iua}, gauge invariant bulk operators should have a representation in terms of boundary operators.  Of particular interest is the study of this mapping for local bulk operators, generally called \emph{reconstruction}, which has been the subject of much study \cite{Banks:1998dd,Hamilton:2005ju,Hamilton:2006az,Kabat:2011rz}, resulting in the famed HKLL procedure, after the authors Hamilton, Kabat, Lifschytz and Lowe. The locality (or microcausality) of these fields should be understood in $G_N$ perturbation theory. In the strict $G_N=0$ limit, these are just free fields in a curved background, but order by order in perturbation theory, gauge invariant bulk fields will have corrections arising from interactions and fluctuations of the background.  We have yet to understand, however, how this approximate notion of locality is consistent with other sensible properties expected of a theory of quantum gravity \cite{Almheiri:2012rt,Almheiri:2013hfa}.

The notion of approximate locality is even less well understood when considering reconstruction in time dependent states. If the state is pure, we expect bulk operators on any Cauchy slice to be  encoded in the boundary. However, this would appear to be violated when considering dynamical boundary states which end up creating event horizons in the bulk. Behind-the-horizon operators should, in principle, be described in terms of their CFT building blocks, but the naive HKLL procedure breaks down in these situations, since the region behind the horizon is not in causal contact with the boundary.

Consider a bulk field $\Phi(X)$ of dimension $\Delta$ in $AdS_{d+1}$. The usual HKLL procedure starts with the extrapolate dictionary, which establishes that near $X=(x,z\approx 0)$
\begin{equation}\label{eq:extrapolate}
  \lim_{z\rightarrow 0}z^{-\Delta}\Phi(x,z)={O}(x,t=0)~,
\end{equation}
Since we are dealing with free fields, the extrapolate dictionary can be extended into the bulk by inverting the massive free field wave equation in a fixed asymptotically AdS bulk. Thus, to leading order in $G_N$,
\begin{equation}\label{eq:HKLL}
  \Phi(x,z)=\int d^{d-1}y\, dt \,K(x,z|y,t) O(y,t)+{\cal O}(G_N)~,
\end{equation}
where $K(x,z|y)$ is the boundary-to-bulk Green's (or smearing) function with support at spacelike separated points from $(x,z)$.

We would like to understand the boundary-to-bulk map in a certain class of dynamical states. The basic idea is to use the equivalence between boundary and bulk time evolution as outlined in \cite{Hubeny:2002dg,Marolf:2008mf,HMPS}. Concretely, consider a local operator $\Phi(X)$ deep in the bulk of empty AdS. For simplicity, we can take $\Phi(X)$ to lie in a Cauchy slice that intersects the boundary at $t=0$. Since black hole horizons are teleological, this operator could well be behind an event horizon depending on if the boundary Hamiltonian does or does not include an injection of energy in the future of the Cauchy slice. The representation of $\Phi(X)$ should be insensitive to the potential future injection of energy, since the bulk operator should not depend acausally on the future evolution. Hence we are free to represent the operator in the case where there is no injection of energy in the future \nref{eq:HKLL}.  To get the representation at later times (and potentially behind the black hole horizon), we simply evolve this operator using the full boundary Hamiltonian, which includes the injected shockwave. The reason this is allowed is that we can think of the right hand side of \eqref{eq:HKLL} as a linear combination of Heisenberg operators at $t=0$. So far, we have been schematic about the precise definitions of $X$ (in particular $z$) and $K(X|y)$, as well as their dependence on the semiclassical gravity background in consideration. This will be a crucial part of our discussion. As we will see, giving a precise definition of bulk points for a general class of states is necessary for being able to compare operators evolved by different Hamiltonians. Furthermore, we have not yet discussed the role of $\mathcal{O}(G_N)$ corrections to \eqref{eq:HKLL}, which will also be crucial for the aforementioned reason.

In summary, we are going to focus on reconstruction in dynamical backgrounds, specifically the thermal state with shockwave insertions on top of it.  We will examine the canonical algebra of local bulk operators generated by $\{\Phi(X),\Pi(X)|X\in \Sigma\}$ in a precise way and understand their boundary representation. Since these bulk operators don't depend on the data away from their respective Cauchy slice \cite{HMPS}, we can think of them as living in an auxiliary, simpler, spacetime. This will not be enough to reconstruct them, but combining this fact with possible forwards and backwards evolution using distinct Hamiltonians, one can reconstruct bulk operators beyond the causal horizon. This procedure will yield an expression similar to \nref{eq:HKLL}, where the boundary operator is a combination of single trace Heisenberg operators, evolved with respect to different Hamiltonians. For simple examples, we show that this expression is equivalent to modular flow, as argued in \cite{Jafferis:2015del,Faulkner:2017vdd}. Such expressions are not unique, and to compare different possible representations of the bulk field, one has to be careful in keeping track of the ${\cal O}(G_N)$ corrections. We illustrate some of the subtleties in simple AdS$_2$ and AdS$_3$ examples. Our procedure is, at its outset, Lorentzian and doesn't rely on analytic continuation to Euclidean times as has been discussed before in \cite{Kraus:2002iv,Hamilton:2006fh,Papadodimas:2012aq,Papadodimas:2013jku}.

In section \ref{sec:review}, we will review and expand on the ideas necessary for reconstruction in the dynamical states of interest. In section \ref{sec:shock}, we define the dynamical states under consideration and the algebras of operators in those states. In section \ref{sec:recon}, we explain how to obtain a boundary expression for operators in any bulk region, as well as how these expressions are compatible with modular evolution. In section \ref{sec:explicit}, we consider the simple examples of AdS$_2$ and AdS$_3$, where the subtleties of the previous sections can be explored and understood explicitly. We conclude with a discussion in section \ref{sec:discussion}.

\section{Toolkit for reconstruction}
\label{sec:review}

 In this section, we will discuss some of the necessary tools needed to deal with the problem of bulk reconstruction in simple time dependent backgrounds. Parts of this discussion have appeared in the literature in various places, and we add new observations to this body of work.

\subsection{HMPS reconstruction}\label{sec:HMPS}

Bulk reconstruction using the HKLL prescription can be implemented order by order in $G_N$ and works for arbitrary asymptotically AdS backgrounds, including ones that are time dependent.
An important refinement of the HKLL story was presented in \cite{HMPS}, whereby a general set of constraints were derived for consistency of the boundary to bulk map (see also \cite{Hubeny:2002dg,Marolf:2008mf} for earlier discussions). We will use these constraints to frame our discussion. The main takeaways from \cite{HMPS} that we would like to highlight are the following:
\begin{enumerate}
  \item Operators $\Phi(X)$ defined on some Cauchy surface $\Sigma_t$ are independent of  the semiclassical background metric.\label{pt:bgkindep}
  \item The bulk Hamiltonian is a boundary term. This implies that bulk and boundary time evolution are equivalent. The Heisenberg evolution of the boundary operator in the RHS of  \nref{eq:HKLL} should match the bulk evolution of the LHS. Namely, this implies that the Heisenberg evolution of an operator using the boundary Hamiltonian should match the representation obtained from applying the HKLL prescription in the time-dependent background.\label{pt:bdryevo}
  \item Operators $\Phi(X)$ on $\Sigma_t$ (understood as Heisenberg operators at fixed boundary time $t$) only depend on the boundary Hamiltonian $H(t)$ in a small neighborhood of the boundary time $t$~.

   \label{pt:localint}

\end{enumerate}

Point \ref{pt:localint} is a statement about causality. It is the claim that for any time $t $, $\Phi(X)$ defined in $\Sigma_t$ will have the same representation in terms of boundary operators ${O}$, independent of the Hamiltonian at $t'>t$. This must be the case if causality in the boundary is to result in a notion of causality in the bulk.

For the sake of illustration we now consider
evolving a state with two different boundary Hamiltonians $\Hold$ and $H(t)$, with $H(t<0)=\Hold$. 
 For the remainder of this paper we will always take $\Hold$ to be a time independent Hamiltonian whose ground state is dual to empty AdS, and label time slices in this empty and static AdS by $\told$. A simple example of $H(t)$ would be the original Hamiltonian $\Hold$ deformed by a source with support localized at $t=0$ such that the bulk dual is the Vaidya collapsing black hole.
With this in mind points \ref{pt:bgkindep} and \ref{pt:bdryevo} taken together suggest the following relation:
\begin{equation}\label{eq:newvsold}
U^{\dagger}_{H(t)}(t',0) \Phi(X) U_{H(t)}(t',0)=U^{\dagger}_{\Hold}(\told',0) \Phi'(X) U_{\Hold}(\told',0)
\end{equation}
where $X$ is a fixed point in the time-independent vacuum AdS geometry. We will present a careful definition of the bulk point $X$ in the next section once we discuss gravitational dressing, such that $X$ can be defined across the families of geometries we are considering. This equation is a bit mysterious at this point as not all the symbols have been sufficiently defined.

In order to understand this equation when we expand $\Phi(X)$ in terms of boundary operators, all $G_N$ corrections must be to be taken into account. To be precise, the leading order HKLL expression for $\Phi(X)$, as in \eqref{eq:HKLL}, seems to depend heavily on the semiclassical background, hence the evolution by different Hamiltonians giving rise to the same operators can only be manifest once the background dependence is removed, e.g. by summing all $G_N$ corrections to the HKLL formula in (\ref{eq:HKLL}). In what follows we will make much use of (\ref{eq:newvsold}) in defining an algorithm for reconstructing bulk operators in terms of CFT data in time dependent shockwave geometries.

We now return to the subtleties in defining the bulk point $X$ in a gauge invariant way---this is particularly pressing, as we will see, for time-dependent backgrounds. We need to supplement the local bulk field $\Phi(X)$ with a ``gravitational dressing'' in order to define it as a gauge invariant operator in the bulk. Treating the gravitational dressing carefully will lead to important considerations for reconstruction in time dependent backgrounds.

\subsection{Gravitational Dressing}

Physical observables $\Phi(X)$ in a theory of gravity must be invariant with respect to bulk diffeomorphisms. In the presence of a boundary, a bulk operator can't commute with the boundary Hamiltonian because of the gravitational Gauss law. These two points imply that $\Phi(X)$ can not be a local operator: $X$ has to be defined in a coordinate invariant manner and $\Phi(X)$ has to include its own gravitational field. However, these operator often still have compact support in the bulk, which provides a notion of locality:
\begin{equation}
  [\Phi(X),\Phi(X')]\approx 0
\end{equation}
 when the non-local operators $\Phi(X),\Phi(X')$  are spacelike separated.\footnote{That is, not only are $X, X'$ spacelike separated so are their respective gravitational fields.}

The operator $\Phi(X)$ is not local, but can be constructed out of a non-diffeomorphism invariant local operator by supplementing it with a ``gravitational dressing.'' This is similar to the story in electromagnetism, where one attaches a Wilson line to charged matter fields to obtain gauge invariant operators. 
 In the context of gravity,  we should think of the dressing as conjugating the matter fields by a unitary (as discussed in \cite{Donnelly:2015hta}): $\Phi(X)=e^{i P v} \Phi(x_0) e^{-i P v}=\Phi(x_0+v)$, where $x_0$ is a point in the bulk geometry and $x_0+v$ provides a gauge invariant characterization of this point, for example, by shooting geodesics towards it from the boundary. $W_\Gamma=e^{i P v}$ is thus the gravitational analogue of the Wilson line and $v$ is a functional of the background metric.

We would like to consider dressing the bulk operators with geodesics. This is a convenient choice of dressing because its support in the bulk is only localized along the geodesic; but more non-local choices of dressing might be natural from the point of view of the boundary theory \cite{Lewkowycz:2016ukf}.  While it would appear that keeping track of the gravitational dressing makes things technically difficult, it is actually not too hard to do so in practice. As was explained in \cite{Donnelly:2015hta}, a convenient way to take the dressing into account is to fix the right gauge. Given a choice of geodesic dressing, \cite{Donnelly:2015hta} showed that if one works in coordinates labeled by the geodesics of the dressing in consideration, the contribution from the dressing disappears $W_{\Gamma}|_{\Gamma-\text{gauge}}=1$. Of course, the commutation relations stay the same and in order to account for the non-locality of the operators, one has to carefully add the additional constraints arising from gauge fixing to the proper Dirac bracket calculation. We refer the readers to \cite{Donnelly:2015hta} for more details.

The simplest example of geodesic dressing is that of perpendicular geodesics. This is usually done \cite{Heemskerk:2012np,Kabat:2013wga} by defining the point $X=(x,z)$ as obtained by shooting a spacelike geodesic perpendicular to the boundary at $x$ and defining
\begin{equation}
  \log\,z\equiv\int_0^1 d\lambda\sqrt{g_{\mu\nu}[y(\lambda)]\frac{dy^\mu}{d\lambda}\frac{dy^\mu}{d\lambda}}~,
\end{equation}
appropriately regulated due to the boundary being at infinite proper distance from points in the bulk. One can show that this is equivalent to fixing Fefferman-Graham gauge for the metric. However, as was explained in \cite{Jafferis:2017tiu}, one can't always define these operators in arbitrary background metrics.

\begin{figure}
\begin{center}
\includegraphics[height=5cm]{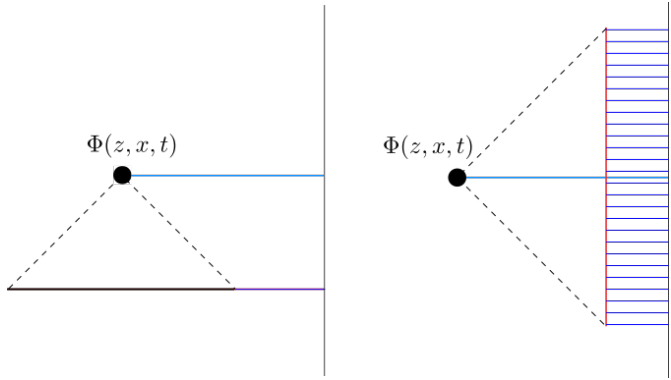}
\caption{A bulk local operator expressed using a retarded(left)/spacelike(right) Green's function. The green line denotes the gravitational dressing of the operator. This operator, along with its smearing can be represented in terms of initial data for the matter part (red smearing) which itself is graviationally dressed (blue).}\label{GravitationalDressing}
\end{center}
\end{figure}

In what follows, we will think of dressing and gauge-fixing interchangeably. What we hope to convey is that, given a choice of geodesic dressing, one can always fix a gauge where the dressing is invisible.\footnote{This is true so long as the respective labeling of points is non-singular.} And with this gauge choice it is very easy to keep track of the dressing. For example, if we have a bulk operator at some time $t$ and we write it in terms of operators at an earlier time using the retarded Green's function :
\begin{equation}
\Phi(z,x,t)=\int_{\Sigma_0} \sqrt{h} dx' dz' (G_{ret}(z,x;z',x')\partial_{t} \Phi(z',x') - \partial_{t}G_{ret}(z,x;z',x')\Phi(z',x'))
\end{equation}
then the dressing can be easily restored from the operator dependence in the left and right hand side: the operator dressed to $(x,t)$ in the boundary is written in terms of operators dressed at other boundary points $(x',0)$ (see the left hand side figure \ref{GravitationalDressing}).  In a similar vein, HKLL reconstruction using a spacelike Green's function---which maps a bulk operator at one point to a linear combination of operators along a timelike surface closer to the boundary--- can be understood as mapping a geodesically dressed operator to a linear combination of operators each of which is dressed to a different point (see the right hand side figure \ref{GravitationalDressing}).  Of course since this is an operator identity, the linear combination of operators is equivalent to the geodesically dressed bulk operator.

\subsection{Background independence and resummation}\label{sec:resum}

Let us further elaborate on point \ref{pt:bgkindep}. Across the paper, we will focus on the the semiclassical regime, where the metric is treated classically. We will consider quantum scalar fields on top of this geometry whose backreaction is small.  To suppress graviton loops we will work to leading order in $G_N$ (but gravitons could still be present as free matter). We also will not consider the backreaction of matter fields.\footnote{The simplest setup in which one can account for  backreaction would be to consider a set of scalar fields with a large number of flavors, $N_F=\alpha N^2 \gg 1, \alpha = \mathcal{O}(1)$. In this limit, the gravitational loops are still suppressed and the backreaction will only come from the scalar fields \cite{Callan:1992rs}.} All of this is in order to only considering boundary states which differ from each other in that their duals have different classical metrics. Since we are ignoring the backreaction from matter, these different states will solve the gravitational Einstein equations in the presence of boundary values for the stress tensor whose expectation value is large, i.e. $\langle T \rangle \sim 1/G_N$.

Our setup will be that of a scalar field in a classical background. The explicit HKLL expressions for the bulk operators appear to depend on the state through the kernel. That is, the kernel is obtained by solving the wave equation in a fixed background:
\begin{equation}
(\nabla^2_{g_0}+m^2) \Phi(X,t)=0\quad\quad \leftrightarrow\quad\quad \Phi(X,t)=\int d^{d-1} x\, dt' K_{g_0}(X,t|x,t') O(x,t')~.
\end{equation}
However, as discussed in \cite{HMPS} the bulk operator $\Phi$ should be independent of the particular value of background field $g_0$. This is no different from electromagnetism: turning on a background electric field changes the wave equation, but this does not imply that the respective gauge invariant operator $\Phi(X,t)$ depends on the background field.

The same should hold in gravity, but it is more subtle, since we are now dealing with different background geometries and one has to carefully define the location and meaning of bulk points (as described in the previous subsection). Gauge invariant ``local" operators $\Phi(X,t)$ are defined in terms of some affine distance along a geodesic thrown from a particular point in the boundary. In order to consider the operator $\Phi(X,t)$ in different classical geometries, one has to make sure that the same geodesic is still contained in the Wheeler-de-Witt patch which corresponds to the boundary time $t$. In other words, one has to require that the new geometry has the ``same" geodesic, as explained in \cite{Jafferis:2017tiu}. These considerations are enough for defining a particular operator at any given point $X$, but not for describing all operators in a Cauchy slice $\Sigma_t$.

However, one might also want to define the algebra of gauge invariant operators in a Cauchy slice $\Sigma_t$. To do this, one has to make sure that one can label all points in $\Sigma_t$ uniquely by $(X,t)$. This requires, among other things, that these geodesics don't have caustics.\footnote{Since we only focus on a single Cauchy slice, it would suffice if they have no caustics at time $t$.} This is a stronger constraint and is equivalent to saying that $(X,t)$ is a well defined set of coordinates in a certain class of geometries. If a family of geometries $\lbrace g \rbrace$ allows for this nice labeling of points, we  define the background independent operator $\Phi_{\lbrace g \rbrace}(X)$ to be the corresponding low energy bulk operator when acting in any member of $\lbrace g \rbrace$.
As an example, consider the family of Ba\~{n}ados geometries \cite{Banados:1998gg}, which can all be written in Fefferman-Graham gauge. In this gauge, $X$ is defined for all members of the Ba\~{n}ados geometries as $X=(x,z)$. In the next subsection, we expand on this example.

In our setup the metric can be treated as a background field. We can then think of the background independent bulk operator as a bulk field that satisfies the wave equations for an arbitrary metric in a particular the family of metrics $\hat{g}$ :\footnote{Since everything in our discussion is gauge invariant, we should think of $\langle \hat{g}\rangle \in \lbrace g \rbrace$  as a function of the stress tensor $\hat{T}$.}
\begin{align}
\left(\nabla^2_{\hat{g}}+m^2\right) \Phi(X,t)=0\quad\quad \leftrightarrow\quad\quad \Phi(X,t)&= \int d^{d-1} x dt' K_{\hat g}(X,t|x,t') O(x,t')~,\nonumber\\ \langle K_{\hat g} \rangle_g&= K_g~.
\label{eq:bindepEOM}\end{align}
The operator $\Phi(X,t)$ is background independent, but has a different form when evaluated in different states.

Given these definitions, we can illustrate what is often called resummation. Consider a particular background metric, $g_0$ and a one parameter family of metrics $g_{\lambda}$, such that $g_{\lambda} \in \lbrace g \rbrace$. For any given $\lambda$, we can schematically expand $K_{g_{\lambda}}$ in $\lambda$: $K_{g_{\lambda}}=K_{g_0}+\lambda K'_{g_0} \partial_{\lambda} g_0+...$ (we are suppresing indices).   This is often interpreted as solving the wave equation order by order in the fluctuations of the metric
\begin{align}
\left(\nabla^2_{g_0}+m^2\right) \Phi_{\lambda}(X,t) & =-\nabla^2_{g_0+\lambda \delta g}|_{{\cal O}(\lambda)}  \Phi_{\lambda=0}(X,t)+{\cal O}(\lambda^2)~, \nonumber \\  \Phi_{\lambda}(X,t) & = \int d^{d-1} x\, dt' \left\lbrace K_{g_0}(X,t|x,t')+\lambda K'_{g_0}(X,t|x,t') \partial_{\lambda} g+\dots\right\rbrace O(x,t') \label{EOMpert}
\end{align}
and then resumming all contributions to get the answer at finite $\lambda$. Note that when evaluated in states outside the family $\{g\}$, the point $X$ so defined won't have a simple meaning. It could be that this point does not even exist in the state in consideration or that the same physical point could be labeled by two different $X$. Let us reiterate: in order for all points $X$ to have a nice interpretation in a family of geometries, one needs to be able to fix the same gauge for all the metrics in this family.

Before we end this section we would like to make some clarifications. Fistly, in the literature, resummation and ``all orders in $G_N$'' are terms often used indistinguishably. However, in the way that we set it up, we want to make a distinction. When solving the equations of motion to all orders in $G_N$, one has to account for graviton loops and backreaction to formally write the bulk operator in terms of the boundary stress tensor $\hat{T}$ (and other boundary operators). The expansion of the bulk field around a different classical background is equivalent to a large shift of the stress tensor $\hat{T} \rightarrow \langle T \rangle+\hat{T},$ with  $\langle T \rangle \sim {\cal O}(N^2)$ in the HKLL expansion to all orders in $G_N$. Only a small subset of the $G_N$ corrections gets enhanced by this shift. Terms stemming from graviton loops or matter backreaction stay of the same order (as they are not too sensitive to one point functions)  and thus, the diagrams that contribute to resummation are due to the shift in the expectation value of the metric in the scalar wave equation \nref{eq:bindepEOM}. This is why we chose to illustrate the notion of resummation in terms of a scalar field in a classical background.

 Secondly, we note that for the same family of metrics $\lbrace g \rbrace$, there might be multiple operators (for illustration we can think of just two) which live in the same geodesic in the metric $g_0$, but live along distinct geodesics for some other metric $g_{\lambda}$. If we denote $\Phi_{\eta}(\lambda,x)$ as the gauge invariant operator at some distance $\lambda$ along a geodesic which has a boost $\eta$ with respect to the boundary, we could consider these two characterizations of the operator: one whose geodesic is always the ``same" in each metric: $\Phi_{0}(\lambda,x)$ (perpendicular to the boundary), and another whose geodesic boost angle depends on the background in consideration: $\Phi_{\alpha f(\hat{g})}(\lambda,x)$,such that it has zero boost angle with respect to the boundary when evaluated in the original background, $f(g_0)=0$. These two operators, when expanded around the original background give the same leading in $G_N$ correlators, but their difference can be diagnosed when acting on other states (or in higher order correlators). As we have explained before, picking a dressing can be easily tracked by fixing a gauge. In our discussion, our gauge choice needs to be defined for a whole family of geometries. In this language the gauge will clearly be different for the two operators we just described. These two operators are clearly state independent and present the usual background dependence as discussed for example in \cite{Harlow:2014yoa}.
\subsubsection{Resummation and large diffeomorphisms}\label{sec:banadosresum}

Let us elaborate on our previous discussion of resummation in the context of AdS$_3$ Ba\~{n}ados geometries. The content of this section has been discussed previously in \cite{Guica:2016pid}, with a few distinctions.\footnote{v2 of \cite{Guica:2016pid} has a mistake to be corrected in v3.}  Consider the Ba\~{n}ados geometries \cite{Banados:1998gg}:
\begin{eqnarray}
ds^2=\frac{dz^2-dx_{+} dx_{-}}{z^2}+T_{+}(x_{+}) dx_{+}^2+ T_{-}(x_{-})  dx_{-}^2- z^2\,{T_+ T_-}\, dx_{+} dx_{-} 
\end{eqnarray}
where Fefferman-Graham gauge has been fixed for all metrics in this family. In this way, the natural geodesic dressed operator are labeled by $X=(z,x_+,x_-)$.  We can consider the HKLL operator, expanded around a fixed background:
\begin{equation}
\Phi_{T_+,T_-}(z,{x_+},{x_-})=\int d{y_+} d{y_-}\, K_{T_+ T_-}(z,{x_+},{x_-}|y_+,y_-) O({y_+},{y_-})~.
\end{equation}
In this case, resummation can be expressed in simple terms since these geometries are related to the to the vacuum AdS$_{3}$ solution:
\begin{equation}\label{eq:bangroundstate}
  ds^2=\frac{d\tilde{z}^2-d\tilde{x}_+ d\tilde{x}_{-}}{\tilde{z}^2}
\end{equation}
by the following large diffeomorphism  \cite{Roberts:2012aq}:
\begin{equation}\label{eq:banresum}
\tilde{z}=z \frac{(f_+'(x_+) f_-'(x_-))^{3/2}}{f_+'(x_+) f_-'(x_-)-\frac{z^2}{2} f_+''(x_+) f_-''(x_-)},\quad\quad \tilde{x}_{\pm}=f(x_{\pm})+\frac{z^2}{2}\frac{ (f_\pm')^2 f_\mp''}{f_+' f_-'-\frac{z^2}{4} f_+'' f_-''}~,
\end{equation}
where $T_\pm$ is related to $f_\pm$ via the Schwarzian derivative:
\begin{equation}
T_{\pm}=-\frac{1}{2}\lbrace f_{\pm},x_{\pm}\rbrace=\frac{3f_\pm''^2-2f_\pm'f_\pm'''}{4f_\pm'^2}~.
\end{equation}

Given that we have a gauge invariant bulk operator, the operator will be the same independent of what coordinates we use to label the point $X$, so we necessarily have that\footnote{One can check that even if one has to evolve the spacelike Green's function up to a different cutoff $y=\epsilon \rightarrow \tilde{y}=\tilde{\epsilon}=\epsilon f'_+(x_+) f'_-(x_-)$, this doesn't really change the RHS.}
\begin{equation}
\Phi_{T_+,T_-}(z,{x_+},{x_-})=\Phi(\tilde{z},\tilde{x}_+,\tilde{x}_-)=\int d\tilde{y}_+ d\tilde{y}_-\, K_{0}(\tilde{z},\tilde{x}_+,\tilde{x}_-|\tilde{y}_+,\tilde{y}_-) O(\tilde{y}_+,\tilde{y}_-)  \label{banadoshkll}
\end{equation}
We should think of the RHS as a function of $f_{\pm}(T_{\pm})$, with $T_{\pm}$ an operator which gets different expectation values (we can equivalently think of effectively promoting $f_{\pm}$ to an operator as in \cite{Turiaci:2016cvo}). When $T_{\pm}=0$, the tilded coordinates are just the original coordinates (up to an $SL(2,\mathbbmsl{R})$ transformation) and this is the usual HKLL expression. For finite $T_\pm$, the HKLL expression gets corrections from the transformation $\tilde{X}(X,T_{\pm})$, implying that we should view the RHS as containing all the corrections due to the shift $T_{\pm}=0 \rightarrow T_{\pm}$, which, in this case these, are just the large diffeormorphisms.

 Note that the kernel $K_0$ transforms as a shadow operator of dimension $(1-h,1-h)$ under conformal transformations of the $\tilde{y}$ argument,\footnote{For global conformal transformations of linear combinations of HKLL operator this was discussed in \cite{Czech:2016xec,daCunha:2016crm}. It can be shown to be true from the expression for the HKLL kernel found in \cite{Faulkner:2017vdd}.} and thus in the previous expression we can send $\tilde{y} \rightarrow \tilde{y}(y)$, since the Jacobian factors from the transformations cancel.

In order to check the extrapolate dictionary, one has to account for the fact that the surface $z=\epsilon$ correspond to $\tilde{\epsilon}=\epsilon (f'_+ f'_-)^{1/2}$. In this way, the $z \rightarrow \epsilon$ limit of the bulk operator gives
\begin{eqnarray}
\epsilon^{2 h} O({x_+},{x_-})=\Phi_{T_+,T_-}(\epsilon,{x_+},{x_-})=\Phi(\tilde{\epsilon},\tilde{x}_+,\tilde{x}_-)=\epsilon^{2 h} (f'_+ f'_-)^h O(\tilde{x}_+,\tilde{x}_-) \label{extrapolatebanados}
\end{eqnarray}
So, the extrapolate dictionary is recovered in the $\tilde{x}$ coordinates after accounting for the transformation of the boundary field under $f$.

\section{Shockwave geometries, geodesically dressed bulk operators and time evolution}\label{sec:shock}
In this section, we describe the class of states to be considered. We will focus on the family of holographic states generated by acting with unitary deformations on the thermal state $\rho_{\beta}=e^{-\beta \Hold}$, that is holographic states of the form $\rho_U= U e^{-\beta \Hold} U^{\dagger}$. These include dynamical processes such as adding sources or more generally evolution with a time dependent Hamiltonian $U(t)=T e^{i \int^t dt' H(t')}$. Note that this includes the case where the ``seed" state is the vacuum, which can be thought as the $\beta \rightarrow \infty$ limit of our discussion. While parts of the discussion will be more general, we will often think  of the states $\rho_U$ as describing the insertion and absorption of shockwaves from the boundary theory, so we will think of these states as shockwave states and for simplicity we will restrict to translation invariant\footnote{Or spherically symmetric if we work with $R_{t} \times S^{d-1}$ on the boundary.} states.

The original ``seed'' state $\rho_{\beta}$ is invariant under time evolution with the Hamiltonian $\Hold$. It will be useful to distinguish between the time generated by the underformed time independent Hamiltonian $\Hold$, $\told,$ and the new time dependent Hamiltonian $H(t)$, $t$. Correspondingly, operators in the interaction picture (i.e. time evolved with the undeformed Hamiltonian) will be $O(\told) = e^{i \Hold \told} O e^{-i \Hold \told} $, while those in the full Heisenberg picture will be $O(t) = U(t) O U^\dagger(t)$.

The class of states generated via conjugation with time dependent Hamiltonians all have the same Von Neumann, or entanglement, entropy as they are all unitarily related to one another. The original state $\rho_\beta$ is dual to the exterior of a black hole with inverse temperature $\beta$ and whose entropy, to leading order in $1/G_N$, is just given by the area of the black hole horizon. This bifurcate horizon is the Ryu-Takayanagi (RT) surface of the entire CFT in the state $\rho_\beta$. The holographic interpretation of this uniformity of the entropy in this class of states is that the effect of the deformations is always causally disconnected from the horizon or RT surface \cite{Headrick:2014cta,Wall:2012uf}. This uniformity of the geometry near the horizon for the entire class is a defining feature of this set of states. As we will later show, the RT surface then serves as an anchor point from which to describe these geometries ``inside out.''

In general, the time dependence of these geometries will cause the RT surface (the bifurcate horizon) to lie behind the new horizon of the now larger black hole. This implies that only a portion of the bulk geometry dual to $\rho_U$ will be causally connected to the boundary. Following the usual terminology from discussions on bulk reconstruction, we will use the term `causal wedge' to denote the region to the exterior of the event horizon,\footnote{Defined by the bulk region which is in causal contact with (can send and receive signals to and from) the boundary.} and the term `entanglement wedge' to denote the rest of the geometry that is bounded by the RT surface and the boundary. The original state has a time translation symmetry which gives a preferred foliation of spacetime, while the other states in this class do not enjoy such a symmetry. However, there are certain Cauchy slices which are natural from the point of view of geodesically dressed bulk operators:  slices foliated by spacelike geodesics.

\subsection{One gauge to rule them all} \label{sec:gauge}
 Given the translation symmetry of the boundary state, the bulk spacetime will be characterized by a set of points consisting of a time-like coordinate $t$,  a space-like ``holographic" coordinate  $Z$ and a boundary space-like coordinate $x$.  In order to talk about gauge invariant operators in this class of geometries, we would like to define the bulk points $(X,t)=(Z,x,t)$ in an equivalent way for all states in our family $\rho_U = U e^{-\beta \Hold} U^{\dagger}$ (at fixed $\beta$). We will define $X$ by shooting geodesics from a fixed reference point,  and therefore $\Phi(X,t)$ will be a geodesically dressed operator.
This can be done by working in so called `geodesic axial' gauge where the metric perturbation along the geodesic is set to zero (that is we set $\delta g_{Z Z}=\delta g_{\mu Z}=0$ where $Z$ labels the proper distance along the geodesic\footnote{We use $Z$ to denote the proper distance along an arbitrary geodesic to distinguish it from the label $z$ often used to describe the proper distance along a perpendicular geodesic.}). This implies that gravitational dressing of the operator will be `invisible' in this gauge.\footnote{This is like having a Wilson line along, say, the $z$ direction and working in the $A_z = 0$ gauge.}

The algebra of scalar field operators is generated by $\{\Phi(X),\Pi(X)| X \in \Sigma_t\}$, where $\Sigma_t$ corresponds to the Cauchy slice dual to the boundary time $t$ (for some foliation). Because we want to consider gauge invariant operators that are geodesically dressed, given a choice of dressing, it seems natural to consider $\Sigma_t$'s that contain all the geodesics that define all $X$.  Given a choice of dressing, this gives a preferred foliation, which we expect to be possible as long as there are no caustics.

However, there are many possible choices of geodesic dressings and in order to single out a particular one, we will use the symmetries of the original state $\rho_{\beta}$. In the original state, we can use the Killing vector to parametrize bulk time and think of this foliation as labeled by throwing constant time geodesics from the RT surface to the boundary.\footnote{Or from the boundary to the RT surface. But, as we will discuss, it is clearer to think of them as being thrown from the RT surface.}
 For an arbitrary state within the class $\beta$, the geometry near the RT surface will have an approximate Killing symmetry. This gives a local preferred foliation, where points are defined by throwing constant Rindler time geodesics from the RT surface.  Because we want our operators to be labeled by geodesics, it seems natural to extend the previous local foliation to the boundary by continuing these (locally constant-in-time) geodesics from the RT surface to the boundary. In other words, we are going to shoot radial space-like geodesics from the RT surface, which we are going to call \ext geodesics. This will give us a foliation of a large part of the space-time\footnote{By shooting geodesics from the horizon there will be a maximum boost after which the geodesic no longer reaches the boundary \cite{Hubeny:2013dea}.} and, for simplicity, we will only consider operators in this region. These Cauchy slices will be labeled by $\told$, the original Killing time near the RT surface, and we will henceforth call them \ext slices. These Cauchy slices have the properties that we expect: they are constant Killing time slices in the original state and for general states they coincide with the Cauchy slices of the original state close to the RT surface. By demanding our operators be geodesically dressed and focusing on Cauchy slices that fully contain the operator and its dressing, this \ext gauge has been singled out.

 This \ext foliation will strongly depend on the details of the state. Consider for simplicity the limit where the geometries consist of compactly supported shockwaves on top of the original state. Away from the shockwaves, the geometry will be that of a static black hole of some given mass. The geodesic emanating from the RT surface will be deflected whenever it crosses a shockwave by an amount related to energy of the shockwave. In this way, a Cauchy slice that corresponds to a particular boundary time, $\Sigma_t$, will depend strongly on all the details of the interior geometry. Going the other way and viewing this foliation as starting from the boundary and going in, due to the translational symmetry, it will be completely characterized by the angle at which it is thrown in. This angle will depend sensitively on how many shockwaves it crosses, so that when it reaches the horizon it corresponds to a constant time geodesic. Therefore the statement is that the mapping  $\told \rightarrow t$ depends sensitively on $U$ used to define the state.

Given this family of geodesically foliated Cauchy slices, the simplest label for bulk points corresponds to the local boost angle at which they are thrown $\told$ and the affine distance $\lambda$ from the RT surface. Because of translation symmetry, we can then label the point from the horizon as $(X_S,\told)=(\lambda,x,\told)$. This labeling is state independent and thus gives us state independent operators; the operator $\Phi(X_S)$ corresponds to the same operator independent of what state in this family we are considering. The $X_S$ label corresponds to dressing the bulk operator to the RT surface.

While these operators are in principle well defined, we might prefer to dress the operator to the boundary. As just emphasized,  $\Sigma_{t}$ will depend sensitively on the background metric dual to $\rho_U$, as it will hit the boundary at a different boundary time $t=t_U(\told)$, depending on the the details or $U$. The affine distance from the RT surface to the boundary $L_{U}(t)$ will also depend on these details as will the angle of incidence of $\Sigma_{t}$ on the boundary. We can thus label the bulk point $(X_S,\told)$ ``outside in'' via $(X_B,t)=(\mathbb{z},x,t)$, where the bulk point corresponds following the respective geodesic along $\Sigma_t$ for a fixed proper distance $\mathbb{z}$. The operators $\Phi(X_B)$ are background dependent: when expanded around a particular background they will lie on the respective geodesics described above. The angle between these geodesics and the boundary will depend on the background, as in the example before section \ref{sec:banadosresum}. When expanded around a particular background, the points $X_S,X_B$ for $\mathbb{z}=L_U(t)-\lambda$ will be the same, but the mapping from $\Phi(X_S) \rightarrow \Phi(X_B)$ is state dependent, since, with $X_S$ we are labeling the operator by its fixed distance to the RT surface while with $X_B$ the operator is labeled by its fixed distance to the boundary.

In this discussion, we are trying to clarify what the meaning of ``the same point'' in this family of geometries should be. This notion of ``same point'' exists because there is part of the geometry which is left invariant by the shockwaves.  That is, as defined $X_S$ and $X_B$ will label the same points if they have the same $x$ as well as the same distance $\lambda=L_U(t)-\mathbb{z}$  from the RT surface. Finally they should land at the same time $\told=\told(t)$ with respect to the near horizon Rindler generator. In this way, the geodesic in the time dependent $\rho_U$ state and the geodesic in the original geometry, dual to $\rho_{\beta}$, respectively labeled by $(X_B,t),(X'_B,\told)$, will have the same endpoint, denoted by $\sim$ as $(X_B,t) \sim (X'_B,\told) \sim (X_S,\told)$ if they are at the same proper distance from the RT surface along the same geodesic. From the boundary point of view, this ``same point'' will be labeled by a different geodesic and proper distance depending on the state.  We will use this characterization heavily in later sections and combine it with time evolution.

There is more than one way to extend the definition of the operators in the original state to other states in the same family, as explained in section \ref{sec:resum}, they will correspond to different operators which have the same leading order in $G_N$ correlators in the reference state. These different definitions correspond to different choices of geodesics which all correspond to the $(z,x,\told)$ geodesic in the original state.

We have just discussed the \ext geodesics, but another natural characterization of the dressed bulk operator would be to always shoot the geodesic from the boundary with zero boost angle. We denote these operators $\Phi(X_{FG},t)$ because they correspond to the Fefferman-Graham (FG) geodesic close to the boundary, which we can extend deeper into the bulk. The point $(X_{FG},t)=(z_{FG},x,t)$ is labeled by the affine distance $z_{FG}$ along a spacelike geodesic from the point $(x,t)$ in the boundary. This definition works very well when one can fix the Fefferman-Graham gauge in the whole geometry as explained in the previous section. However, in the time dependent geometries considered herein, the FG geodesics won't give a natural foliation of the entanglement wedge, since these geodesics generically will not go through the horizon. These geodesics will also have caustics. So, while the $\Phi(X_{FG},t)$ operators can be defined independently of the $\rho_{\beta}$ family in consideration, they don't preserve any notion of locality. As we will explain later, the \ext dressed operators have nicer properties.

\subsection{Algebra of gauge invariant operators}
Consider the algebra of operators $\Phi(X_B)$ which are dressed to the boundary by the \ext geodesics fully contained in the their respective Cauchy slice. The geodesics defining these operators can cross multiple shockwaves and therefore, in order to understand them from the boundary point of view, one has to be careful about their dressing.
In order to discuss the algebra of operators at a fixed time, we will consider operators in the Schr\"{o}dinger picture.
 When expanding the Schr\"{o}dinger operators $\Phi(X_B)$ around a given background, they will depend on the background through the time at which they are evaluated (but the operator itself is state independent), this we will denote $\Phi(X_B^t)$.  Thus, the $t$ label is there to keep track of the background and how are we dressing the operator, but the operator itself does  not depend on $t$.

 For clarity, we will denote the Schr\"{o}dinger operator expanded around the original geometry by $\Phi(X_{\Hold})$, since the background dependent boundary condition on the geodesic labeling $\Phi(X_B^\told)$ ends up always being the same in this state. We acknowledge that the background dependence of the dressing is a drawback of the $\Phi(X_B)$ operators in contrast with, for example, $\Phi(X_{FG})$, we still consider the former dressing because it has nicer local properties.

\begin{figure}
\begin{center}
\includegraphics[width=0.6\textwidth]{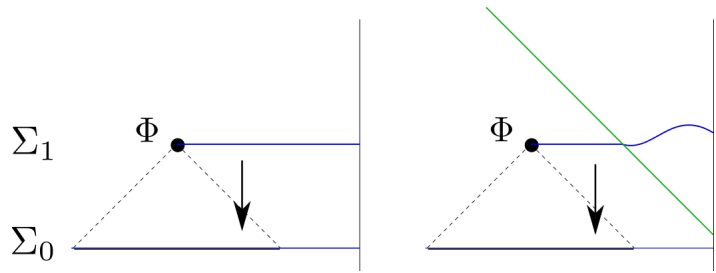}
\caption{Depiction of our definition of ``same point,'' $(X_B^t,t)\sim(X_\Hold,\told)$. Note that the dressings reach the same point in the old geometry as defined by their destance to the original black hole horizon or RT surface. By backwards evolving using the respective Hamiltonians should give equivalent expressions for the operators as described in equation \eqref{eq:newvsoldv2}.}\label{TheEquation}
\end{center}
\end{figure}

Taking everything into consideration, we can write a more precise version of equation \nref{eq:newvsold} while taking the dressing into account:
\begin{equation}\label{eq:newvsoldv2}
U^{\dagger}_{H}(t,0) \Phi(X^{t}_B) U_{H}(t,0)=U^{\dagger}_{\Hold}(\told,0) \Phi(X_{\Hold}) U_{\Hold}(\told,0), ~~ (X_B^t,t) \sim (X_{\Hold},\told)
\end{equation}
where $(X_B^t,t) \sim (X_{\Hold},\told)$ means that the two points are at the same in the sense defined in the previous section.
While $X^t_B$ will cross some number of shockwaves, $X_{\Hold}$ will not. Note that the two points $X_B^t$ and $X_{\Hold}$ will be characterized by different proper distances from the boundary and therefore $\Phi(X^{t}_B)$ and $\Phi(X_{\Hold})$ are not the same operators. See figure \ref{TheEquation}.

What if we had used the FG operators $\Phi(X_{FG})$? Again the ``same point'' when comparing between the time dependent geometry and the dual of $\rho_\beta$ should be understood in terms of having the same geodesic distance to the horizon, i.e. $(X_{FG},t) \sim (X_B,\told)$ means that the two points sit at the same point in the original geometry (and therefore they have different affine distances from the boundary).  However, at this juncture, it is not clear to us if \nref{eq:newvsoldv2} is also true for $\Phi(X_{FG})$, because of the aforementioned issue with caustics.
We explore this further in section \ref{sec:explicitads3}.

Let us reiterate that, as opposed to the FG gauge, the algebra of \ext dressed operators satisfies nice local properties. These geodesics associate one operator per bulk point (since they have no caustics) and $[\Phi(X_B^t),\Pi(X'^t_B)] \propto \delta\left(x-x'\right)$, where, as we described before, $X_B^t=(\mathbb{z},x)$.
We expect that the tradeoff between locality and state dependence (on the family of states)  is generic.
\subsection{Summary of notation}
Let us summarize the notation:
\begin{itemize}
\item $\Phi(X)$ denotes Schr\"{o}dinger operators, which are defined gauge invariantly since $X$ is the endpoint of a geodesic.
\item $(X^{t}_B,t)=(\mathbb{z},x,t)$ is an \ext dressed point which sits at a proper distance $\mathbb{z}$ from the boundary along a geodesic which has some boost angle with respect to the asymptotic boundary but which becomes a constant Rindler time geodesic when it approaches the horizon.
\item $(X_{FG},t)$ denotes a geodesic with zero boost angle with respect to the boundary. Since the boost angle is fixed, the operator defined in this way will always be the same (i.e. background independent as it does not depend on the family of geometries). In time dependent backgrounds these coordinates don't give a nice foliation of the spacetime due to caustics.
\item $(X_{\Hold},\told)$ labels a point in the original geometry $\rho_{\beta}$. Both definitions $X_{FG}$ and $X_{B}^t$ coincide in this state: and correlation functions of these various operators will be the same in the state $\rho_\beta$ to leading order in $G_N$.
\item $(X_S,\told)=(\lambda,x,\told)$ denotes a geodesic with proper distance $\lambda$ from the RT surface, thrown from the point $x$ along the RT surface and at Rindler angle $\told$.
\item We denote the ``same point'' in the bulk across different geometries using $\sim$, as in $(X^t_B,t) \sim (X_{S},\told)\sim (X_{\Hold},\told)$, where  $(X^{t}_B,t)$ can be thought as reaching $(X_{S},\told)$ along the same geodesic but charaterized ``outside-in'' from the boundary. Thus the Schr\"{o}dinger operators $\Phi(X^t_B)$ will depend on the time that they are evaluated, since the boost angle will be different {(but the dependence of the operators is just a background effect)}. One could also try to define some notion ``same point'' for $(X_{FG},t)$, but the identification between $(X_{FG},t)\sim (X_{\Hold},\told)$ will be more complicated.
\end{itemize}

\section{Reconstruction of operators behind shockwaves}
\label{sec:recon}
In this section, we will describe in explicit terms how one can reconstruct operators deep in the entanglement wedge using the fact that the local operators only depend on the boundary Hamiltonian at their respective time (see point \ref{pt:localint}), while being careful about dressing. We will also compare this with the modular flow considerations of \cite{Jafferis:2015del,Faulkner:2017vdd} for reconstructing operators beyond the causal wedge.

\begin{figure}
\begin{center}
\includegraphics[width=0.45\textwidth]{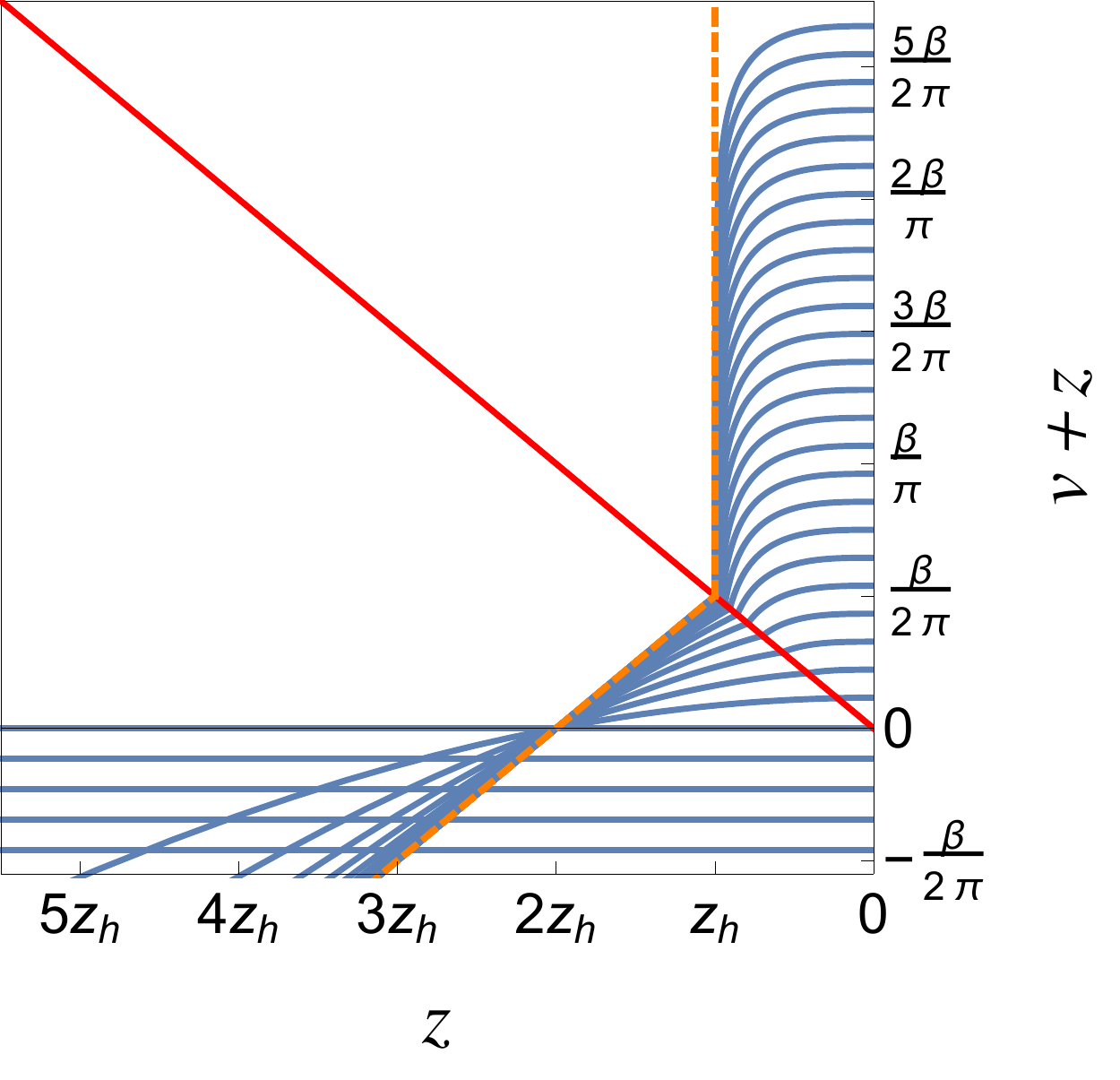}\hfill
\includegraphics[width=0.47\textwidth]{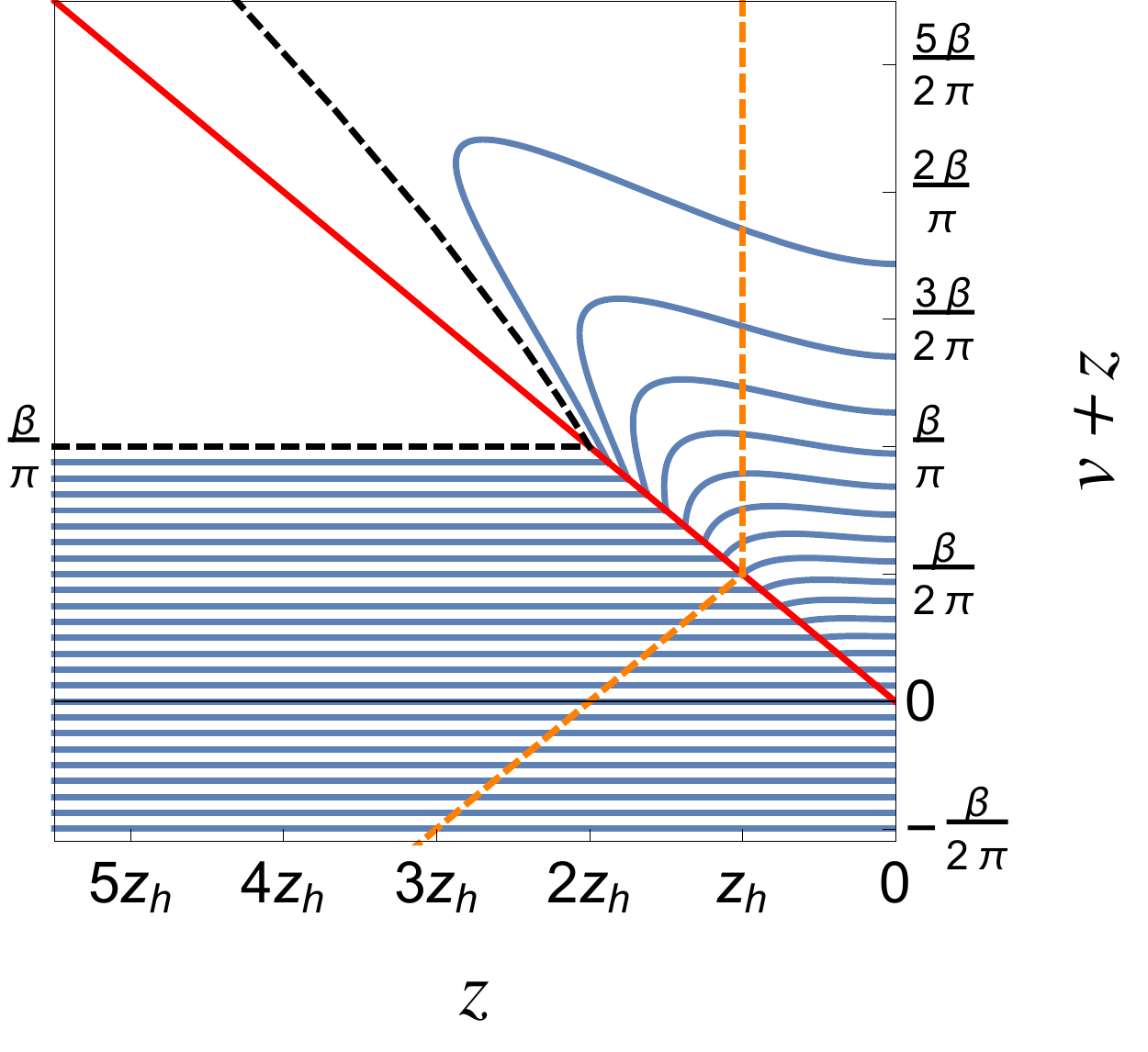}
\caption{\emph{Left}: Fefferman Graham geodesics shot with zero boost angle from the boundary. Note that these geodesics have caustics behind the black hole horizon. \emph{Right}: Extremal geodesics shot with zero boost angle from the RT surface. The black dotted line corresponding to $\told=\beta/\pi$ hits the boundary at $t=\infty$. The orange dotted line is the location of the horizon and the solid red line gives the location of the shockwave. In both cases we plot $(Z(s),V(s)+Z(s))$. T}\label{geos}
\end{center}
\end{figure}

As we mentioned in the introduction, the idea is to use the operators $\Phi(X_B^t)$ and the equality between bulk and boundary evolution. While we will use this set of coordinates, nothing in this section really depends on this specific choice of dressing: while the \ext geodesics that define $\Phi(X_B^t)$ seem to give a nice foliation of the entanglement wedge, we could also consider $\Phi(X_{FG})$ (or other dressings), so the reader can make the substitution without worry, taking into account that when we refer to the ``same point'' in the old and new geometry, we mean it in the sense explained in section \ref{sec:gauge}. At some points in this section, we will appear to draw conclusions that depend on the \ext gauge choice. When this is the case, we will also describe what we expect to happen had we chosen $X_{FG}$.  Note that because of caustics, it is not always clear how to compare  $\Phi(X_{FG})$ at different times (see figure \ref{geos} for an illustration) and that is why we are focusing our analysis around $\Phi(X_B^t)$, which don't suffer from these problems.

As summarized in point \ref{pt:localint}, it was suggested in \cite{HMPS} that the bulk dressed operator only depends on the Hamiltonian at $t$ and we are free to change $H$ elsewhere.  By using the bulk equations of motion and the equality between bulk and boundary evolution, $\Phi(X_B^t)$ can be written in terms of operators at other times:\footnote{After fixing the \ext gauge, the dressing is invisible. Thus we can choose to first fix the gauge to derive this expression using the appropriate wave equation and then write it in a gauge invariant way by adding the proper explicit dressing. }
\begin{equation} \label{ret}
\Phi(X_B^t) =U_{H}(t,t') \int_{\Sigma_{t'}} d\Sigma_{t'} \left ( G_R(X_B^t,Y_B^{t'}) \Pi(Y^{t'}_B)-\partial_{t'} G_R(X_B^t,Y^{t'}_B) \Phi(Y^{t'}_B) \right) U_{H}(t',t)~,
\end{equation}
where we used the boundary unitaries to write this as an operator at $t$. As we we have been keen to stress, even if this expression seems rather simple, since these unitaries $U_H(t,t')$ denote boundary evolution, one needs resummation to see the equivalence between the LHS and RHS in terms of boundary fields.  The reason is clear, the states $\rho_{t'},\rho_{t}$ are macroscopically different and when one uses the \emph{same} HKLL operator in two macroscopically different states one has to resum all the gravitational corrections. In other words the wave equation is different in these two states and therefore the zeroth order HKLL kernels reflect this.

\begin{figure}
\begin{center}
\includegraphics[width=0.55\textwidth]{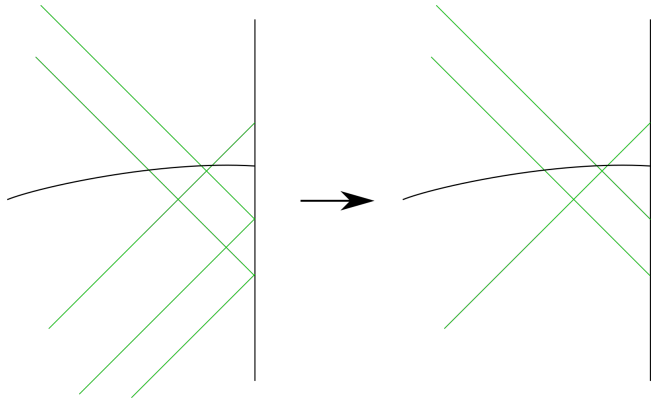}
\caption{Illustration of step \ref{step:remove}, whereby we remove all shocks that do not cross  $\Sigma_{t_0}$, the Cauchy surface our operator lies in.}\label{A}
\end{center}
\end{figure}

\begin{figure}
\begin{center}
\includegraphics[width=0.6\textwidth]{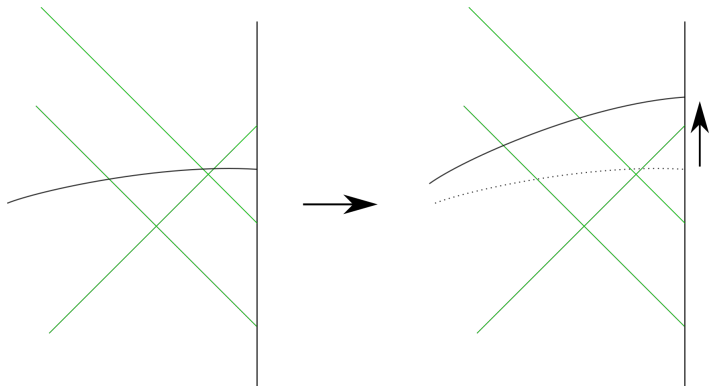}
\caption{Illustration of step \ref{step:move}, whereby we simplify the operator by evolving it to a Cauchy slice that crosses fewer shocks.}\label{B}
\end{center}
\end{figure}

With these caveats in mind, we would like to propose the following procedure for reconstruction in time dependent geometries:
\begin{enumerate}
\item  The Cauchy slice $\Sigma_{t_0}$ where our operators live will generically cross a subset of all the shockwaves present in the spacetime. Given this Cauchy slice, the simplest way to time evolve this state is to consider a Hamiltonian which does not add any new shockwave, other than those which cross the geodesic. For example, if one of the shockwaves is reflected along the boundary at a later time, we will choose a time dependent Hamiltonian which absorbs the shockwave. We will consider evolution by this Hamiltonian $H_{t_0}(t)$. See figure \ref{A} for an example.\label{step:remove}
\item  Given this new geometry generated by the time dependent Hamiltonian $H_{t_0}(t)$, we will use  (\ref{ret}) (or an advanced version of it) to choose a new $t_1$ such that $\Sigma_{t_1}$ crosses a smaller number of shockwaves than the original surface. This is usually done by evolving past shockwave insertions in the boundary.\footnote{One can do this so long as the shockwave we are evolving away from is inserted at a boundary time $t$ which does not scale with $N$. In contrast, the shockwaves in the Shenker-Stanford geometries are inserted far in the past, at a scrambling time $t \sim -N \log N$, and our procedure can not be applied.
}
The goal of this is to simplify the operator. In this procedure,  (\ref{ret}) can in principle include a contribution from fields in the boundary.  See figure \ref{B}.\label{step:move}
\item  Now, we start again with the operators  $\Phi\left(Y_B^{t_1}\right)$, constructed by evolving the original operator using the above steps. Since it does not matter how we prepared the state at time $t_1$, we can just repeat step $1$ for each of these operators independently.\label{step:rinserepeat}
\end{enumerate}

These steps can be repeated until the Cauchy slice where the operator is located does not cross any shockwave and thus the operator can just be reconstructed in the boundary in terms of the usual causal reconstruction in the exterior of black holes.

 Note: it could be that the operator is in causal contact with the boundary before eliminating the last shockwaves. We could use HKLL here. The corresponding operator will of course be equivalent to what we would get by removing all shockwaves. This might seem surprising at first and to see it explicitly, one should solve the wave equation in the presence of gravitational corrections and then shift the background. We discuss this in section \ref{sec:resumtime}.

\begin{figure}
\begin{center}
\includegraphics[width=0.5\textwidth]{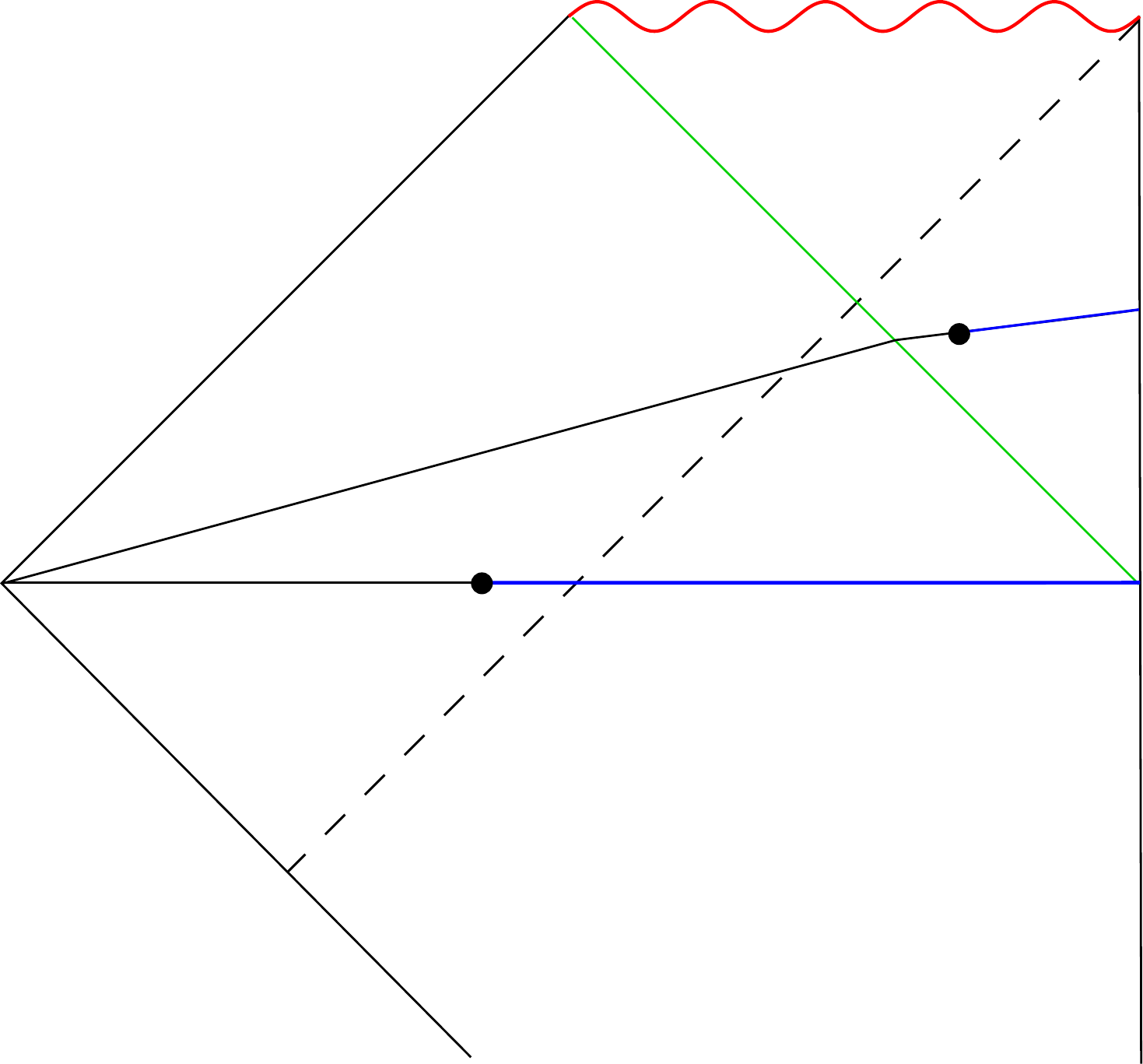}
\caption{Operators in Cauchy slices that either cross a shock or do not. If the operator's dressing crosses no shocks---but the full Cauchy slice does---we can not use step \ref{step:remove} to remove said shock without first using step \ref{step:move} to move the operator to a different Cauchy slice which does not cross the shock.}\label{C}
\end{center}
\end{figure}

It is important to highlight that, in order to remove a shockwave using \ref{step:remove}, we need this shockwave to not cross the entire Cauchy slice $\Sigma_{t_0}$ where our operator lives, at any point. This means that, even if the operator $\Phi(X^{t_0}_B)$ does not cross the shockwave, but the Cauchy slice $\Sigma_{t_0}$ does, we can not change the Hamiltonian so that that shockwave was not inserted. For example, if the Hamiltonian adds a shockwave at some time $t=0$, when reconstructing operators at $t<0$, we can consider the Hamiltonian where there is no shockwave inserted at $t=0$. However, if we want to reconstruct operators at any $t>0$, we can only change the Hamiltonian so that the shockwave is reflected in the boundary, but we can not really get rid of this shockwave (see figure \ref{C} for an illustration). This global constraint avoids possible paradoxes and morally keeps track of the original state $\rho_{\beta}$. If possible we could try to use the retarded Green's function to move $\Phi(X^{t_0}_B)$ to a Cauchy slice which crosses no shocks and then use \ref{step:remove}.

We can illustrate this rather general procedure for Vaidya and double (time symmetric) Vaidya without finding the explicit kernel. In section \ref{sec:explicit}, we will write some explicit expressions in simple examples.

\begin{figure}
\begin{center}
\includegraphics[width=0.6\textwidth]{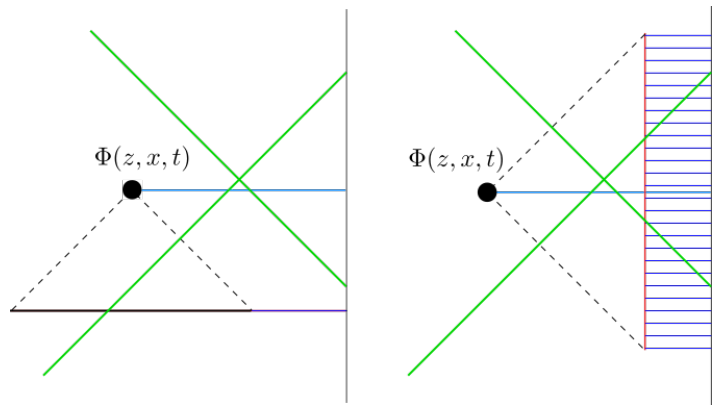}
\caption{We can exploit the spacelike HKLL procedure to write an operator whose dressing crosses multiple shockwaves as a linear combination of operators whose dressing crosses less shockwaves.}\label{Z}
\end{center}
\end{figure}

HKLL can be understood as using a spacelike Green's function to write bulk operators in terms of boundary operators. One could also potentially use the bulk-to-bulk (rather than the usual bulk-to-boundary Green's function used in HKLL) spacelike supported Green's function to simplify some of the previous steps. The idea behind this simplification is that for a bulk operator which crosses two shockwaves which intersect,  we can use this bulk-to-bulk evolution to write the bulk operator in terms of bulk operators closer to the boundary, which only intersect one shockwave. This identity is sometimes helpful deep in the bulk, where the alternative in the outlined procedure would be to time evolve the bulk operator until it only crosses one shockwave (see figure \ref{Z}).

\subsubsection*{Vaidya}

Consider the case of Vaidya:

\begin{equation}
ds^2=\frac{1}{z^2} \left ( -F(z,v) dv^2-2 dv dz+d\vec{x}^2 \right) ~~ f(z)=1-( 2\pi z)^{d} \left (\frac{\Theta(-v)}{\beta^{d}} +\frac{\Theta(v)}{\beta'^{d}} \right) \nonumber
\end{equation}

\begin{figure}
\begin{center}
\includegraphics[width=0.6\textwidth]{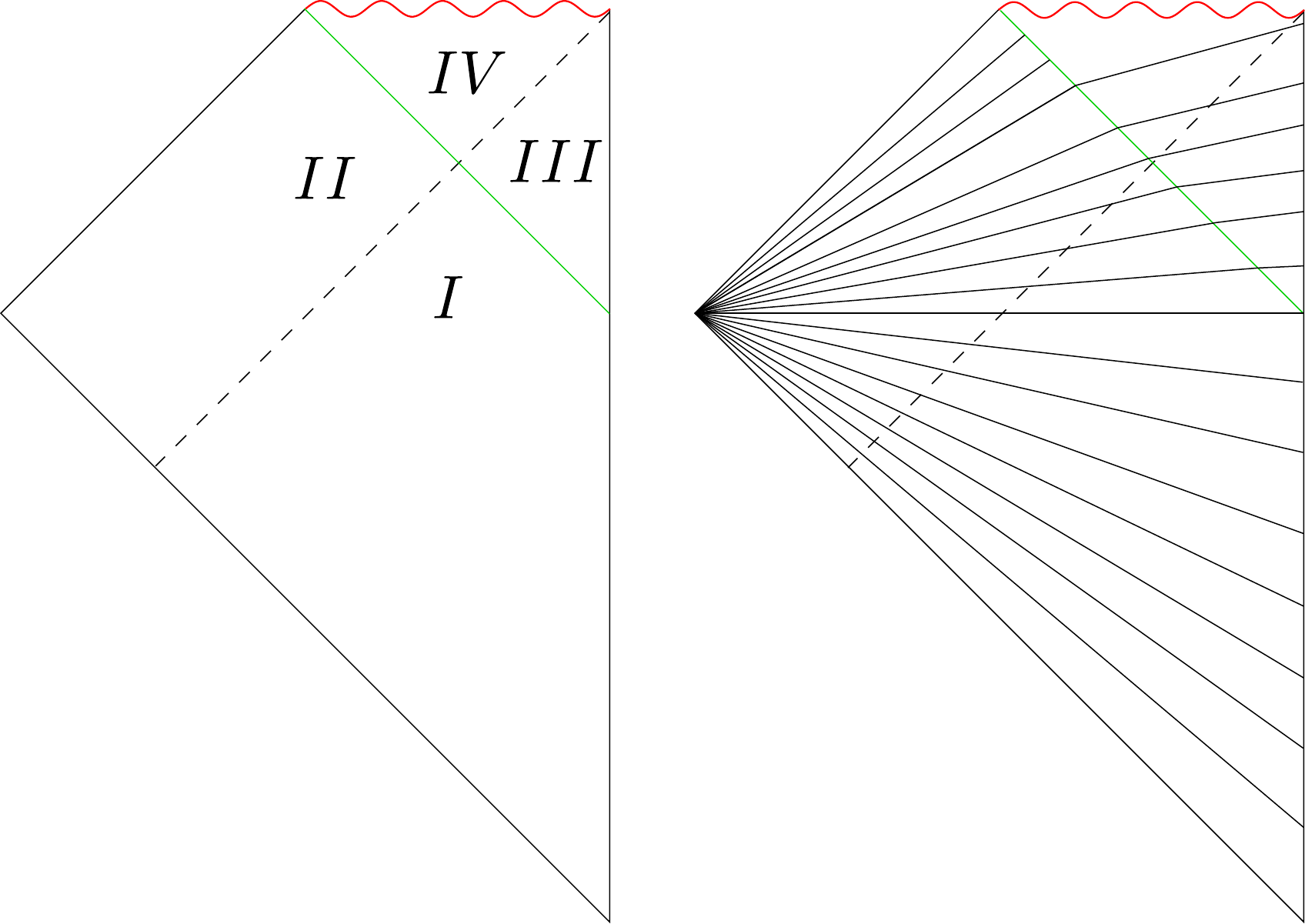}
\caption{Penrose diagram and \ext Cauchy slices for AdS-Vaidya. We label regions of the penrose diagram $I-IV$. There is a maximum boost angle after which the \ext geodesics don't reach the boundary.}\label{E}
\end{center}
\end{figure}

The Penrose diagram for Vaidya is shown on the left diagram of figure \ref{E} and the time slices that we are considering are those in the right diagram. Our discussion thus far has taken the original state $\rho_\beta$ to be a thermal state with temperature $\beta$ and the geometry after the shockwave as that of AdS-Schwarzschild with $\beta'<\beta$. Pure state Vaidya is a special case $\beta \to \infty$ of this general discussion. In the finite temperature case, we are only interested in the entanglement wedge, which is the region outside the original bifurcation horizon (right side of the thermofield double). In the $\beta\to \infty$ limit, the entanglement wedge becomes the whole spacetime.

It is natural to split the Penrose diagram into four regions, depending on whether the points are inside or outside the causal horizon and before or after the shockwave. The regions are labelled I$ -$IV in figure \ref{E}. From the point of view of Cauchy slices and operators, an alternative important division could split the regions depending on whether the Cauchy slice crosses  the shockwave or not. If the shockwave is sent from $t=0$, then the Cauchy slices $\Sigma_{t>0}$ will contain operator which cross the shockwave. Reconstruction in the two regions to the past of the shockwave was discussed using precursors in \cite{HMPS}.

Let us start by discussing the operators in region II with geodesics anchored at boundary times $t<0$. We summarized this case in the introduction. These operators live entirely in the old geometry but they are not in causal contact with the boundary region, so one can not use HKLL reconstruction. Since the Cauchy slices cross no shocks, we can use step \ref{step:remove}, which instructs us to change the Hamiltonian and get rid of the shockwave. This allows us to express the operator in region II in terms of boundary operators evolved with the original, time independent Hamiltonian, $\Hold$. These operators are not in the causal wedge, which implies that they can not be written in terms of simple boundary operators (operators that satisfy the extrapolate dictionary in the shockwave geometry). They can, however, be written in terms of operators evolved with the original Hamiltonian. If the operator in II sits at a time $t>0$, then its Cauchy slice will cross the shockwave. Then, step \ref{step:move} instructs us to use (\ref{ret}) to write the operator in terms of operators whose dressing does not cross any shockwave and we can apply the previous discussion.

Let us now give an equation for the discussion in the previous paragraph. If we denote HKLL in the original state (in the Heisenberg picture) by:
\begin{equation}
U^{\dagger}_{\Hold}(\told,0) \Phi(X_{\Hold})U_{\Hold}(\told,0)=\int d^{d-1} x d\told' K_{\Hold}(X_{\Hold},\told-\told'|x) O(x,\told')+\mathcal{O}(G_N)~,
\end{equation}
where we wrote the kernel $K_\Hold$ in a explicitly time translation invariant form, then we can apply (\ref{eq:newvsoldv2}) to write:
\begin{multline}
U^{\dagger}_{H}(t,0) \Phi(X_B^t)U_{H}(t,0)=\int d^{d-1} x d \told' K_{\Hold}(X_{\Hold},\told(t)-\told'|x) O(x,\told')+\mathcal{O}(G_N),\\\ ~~(X^{t}_B,t) \sim (X_{\Hold},\told)~, \label{HKLLreg2}
\end{multline}
where $\told(t)$ is the near horizon Rindler time which corresponds to the boundary time $t$.\footnote{Recall that by $X_{\Hold}$ we mean the bulk point in the old geometry which has the same proper distance along a constant Rindler time geodesic thrown from the RT surface as compared with $X_B^{t}$. See \ref{sec:gauge} for a reminder.}  Note that equation (\ref{eq:newvsoldv2}) also has unitaries on the RHS, but since the Hamiltonian is time independent, this just amounts to shifting the time argument in the kernel.

Reconstruction in region I is quite similar to the discussion for region II and we can repeat it straightaway. However, since region I is actually in causal contact with the boundary causal domain, we have the option of using HKLL in the Vaidya geometry. For simplicity, let us first consider the operators with $t<0$, lying in a Cauchy slice that does not cross the shockwave. It may seem odd, at first, that the full HKLL expression in the Vaidya background can give the same correlators as the expression given in the RHS of \eqref{HKLLreg2}. This was one of the main points of \cite{HMPS}. The caveat, again, is that the bulk operator is background independent and, thus, in order to compare the two expressions one needs to remember to take resummation into account. That is, one should in principle solve the wave equation for an arbitrary background and then expand around the background in consideration. We will discuss this further in section \ref{sec:resumtime}. In section \ref{sec:explicit}, we illustrate how this works more explicitly, using the symmetries of $2,3$ bulk dimensions.

Region III is morally equivalent to region I. One can use HKLL directly in the new background or evolve to the past and write the bulk operator in terms of boundary operators plus bulk operators in the old geometry. Then one can construct the operators in the old geometry as in I. In this case, it is much simpler and nicer to simply use HKLL in the new geometry.

\begin{figure}
\begin{center}
\includegraphics[width=0.6\textwidth]{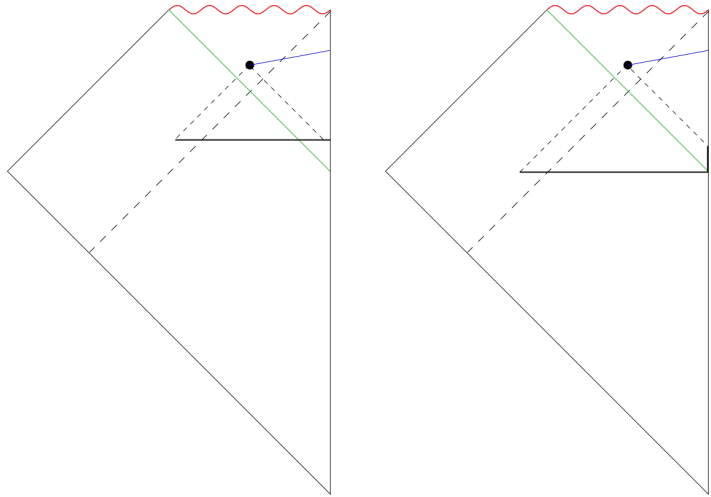}
\caption{Equivalent reconstructions for operators behind the new black hole horizon and to the future of the shock (region IV). \emph{Left}: We can use the retarded Green's function to write $\Phi$ in terms of bulk operators in regions I, II and III. \emph{Right}: We can use the retarded Green's function to evolve to the $t=0$ slice where we can write $\Phi$ in terms of operators in regions I and II as well as boundary operators in the new geometry.}\label{F}
\end{center}
\end{figure}

Region IV is the interior of the new black hole. While in the previous cases, we used step \ref{step:move} to avoid having the dressing cross the shockwave, we now use \ref{step:move} to opposite effect: we write this operator in terms of operators whose dressing now crosses the shockwave (the region to the past of the shockwave) and operators whose dressing does not cross the shockwave, but their respective Cauchy slice does. The operators in region IV can thus be reconstructed in two different but equivalent ways. One can either use the retarded Green's function to write the bulk operator in terms of bulk operators in regions I, II, III and reconstruct the operators in these regions as discussed above (see the left hand side of figure \ref{F}), or one can evolve back to $t=0$ (or $t<0$) and write the bulk operator in terms of bulk operators at $t=0$ as well as boundary operators (see the right hand side of figure \ref{F}).

Note that while the FG dressing is poorly suited to the Vaidya example, it seems that there is another class of natural dressings for this state, which is to dress the operators along ingoing null rays (discussed for example in \cite{Papadodimas:2015jra}). However, using ingoing null rays will not work for more generic geometries made out of shockwaves, such as the reflecting Vaidya geometry discussed in the next section.

\subsubsection*{Reflecting Vaidya}

\begin{figure}
\begin{center}
\includegraphics[width=0.4\textwidth]{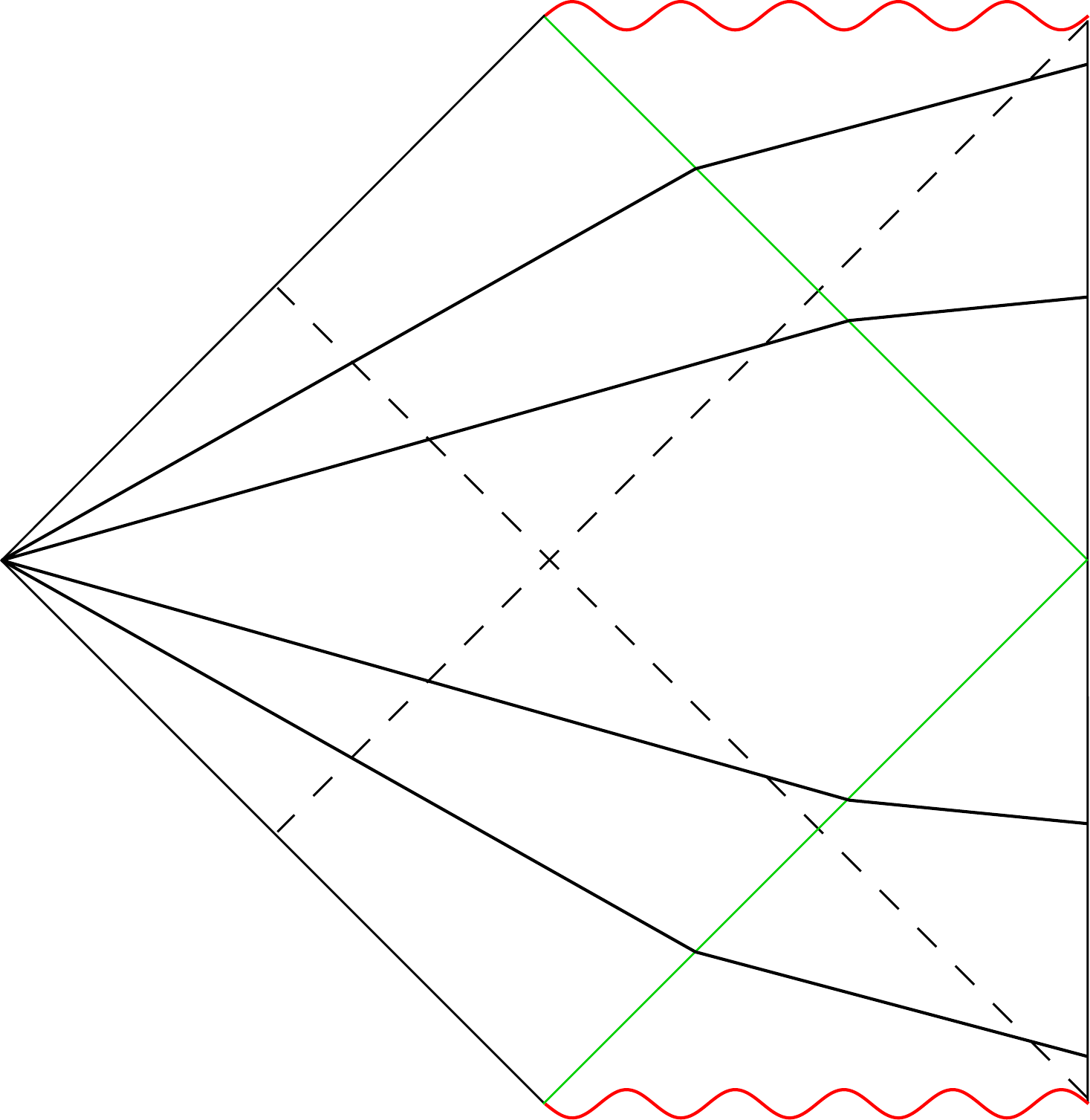}
\caption{Penrose diagram and examples of \ext Cauchy slices for the reflecting Vaidya geometry.}\label{G}
\end{center}
\end{figure}

Consider now a reflecting Vaidya geometry, given by an expanding null shell of matter in the past which collapses again after it reaches the boundary, as in figure \ref{G}. This example is important, because we will show that it is not sufficient to use the almost light-like foliation, useful in the Vaidya case as discussed in \cite{Papadodimas:2015jra}, when there is a past shockwave.  Another crucial difference is that, in this case, all Cauchy slices cross one shockwave.
The state dual to reflecting Vaidya is given by $\rho_W=W \rho_\beta W^{\dagger}$, where $W$ is the shockwave creation operator. The boundary Hamiltonian is none other than the time independent $\Hold$, but the state is not an eigenstate of the Hamiltonian. We will denote the boundary time of the \ext geodesics by $t_W$.

The procedure for reconstructing the operators follows the Vaidya discussion closely. For an operator at $t_W>0$, step \ref{step:remove} instructs us to consider the time dependent Hamiltonian of Vaidya and thus we can then repeat the previous story. For operators at $t_W<0$, the procedure is a time reflection of the previous.

 The limit when $t_W \rightarrow 0$ is of course continuous, so nothing special happens when we consider the operators at $t_W=0$. This $t_W=0$ Cauchy slice is essentially a Cauchy slice in the shock-less state $\rho_\beta$, only differing at the point where it crosses the reflecting shockwave near the boundary. Because of the presence of a shockwave reflecting at infinity, the $\Phi(X_B^{t_W=0})$ operators intersect the shockwave and thus these are different operators from those of the original geometry.  Note that  (\ref{eq:newvsoldv2}) relates the operator at $t_W=0$ in this state with the operator in the state $\rho_\beta$:\footnote{This can be seen by comparing the state $\rho_W$ at some $\Sigma_{0^+}$ with the same state but now evolved with a time dependent Hamiltonian which inserts the shockwave at $t=0$.}
 \begin{equation}
W^{\dagger} \Phi(X_B^{t_W=0}) W=\Phi(X_{\Hold}),\quad\quad\quad X_B^{t_W=0} \sim X_{\Hold}~.
 \end{equation}

\begin{figure}
\begin{center}
\includegraphics[width=0.6\textwidth]{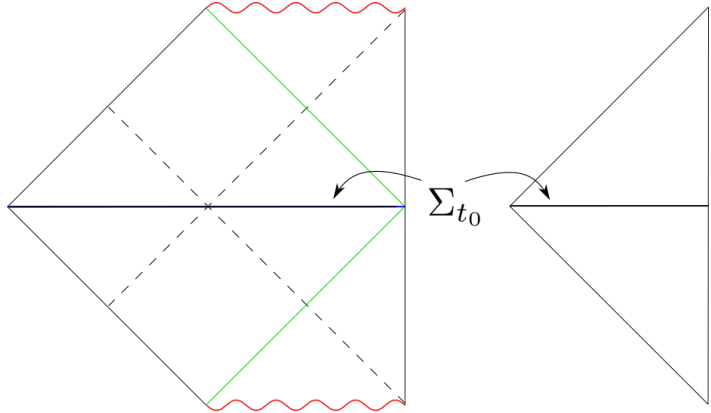}
\caption{The geometry on these slices is identical implying a natural identification of points between them. However, the proper distances from these points to the boundary differs across the two cases because of the shockwave encountered near the boundary.}\label{H}
\end{center}
\end{figure}

The operator in the reflecting Vaidya geometry $\Phi(X_B^{t_W=0})$ is different from the local operator in the original state and this difference is necessary to ensure the algebra of operators at $\Sigma_{t_W=0}$ gives rise to the the same bulk correlators:
 \begin{equation}
 \langle \Phi(X_B^{t_W=0})... \Phi(Y_B^{t_W=0}) \rangle_{\rho_W}= \langle \Phi(X_{\Hold}) ... \Phi(Y_{\Hold})\rangle_{\rho_\beta}~,
 \end{equation}
while, of course, if we consider adding boundary operators into the above correlators there will be no equality. Had we neglected the dressing, we might have concluded that $\Sigma_{t_W=0}$ is essentially the same slice for $\rho_\beta$ and $\rho_W$ and thus the correlators of all operators in this slice should be the same. This certainly can not really be true unless all operators commute with the shockwave. See figure \ref{H} for an illustration.

Note, while it is trivial that an operator $W\Phi(X_{\Hold}) W^{\dagger}$ has the same correlators in the state $W \rho_\beta W^{\dagger}$ than the operators not conjugated by the unitary in the original state, it is non trivial that the operator conjugated by the unitary corresponds to a (properly dressed) bulk local operator at $\Sigma_{t_W=0}$ in the $\rho_W$ geometry.\footnote{For example, if $\Sigma_{t_W=0}$ crossed the shockwave somewhere inside the bulk, the operator $W\Phi(X_{\Hold}) W^{\dagger}$ won't have the interpretation of a local operator in the new geometry if the geodesic that connects $X$ with the boundary does not cross the shockwave, since there is no analogue of this point in the geometry with no shockwave. In this particular example, since $\Sigma_{t_W=0}$ crosses the shockwave near the AdS boundary, $W\Phi(X_{\Hold}) W^{\dagger}$ corresponds to a local bulk operator for all $X \in \Sigma_{t_W=0}$} This can be cast in the language of error correction of \cite{Almheiri:2014lwa}: given the bulk Cauchy slice $\Sigma_{t_W=0}$, which is almost the same for the two states, $\rho_W,\rho_\beta$, the expectation values of the elements in the algebra of low energy operators ${\cal A}_W$ in our state $\rho_W$ does not depend on $W$. That is, we can think of the GNS subspace created by acting with a small number of operators in $\Sigma_{t_W=0}$ and, while the mapping from the bulk to the boundary depends on the explicit details of the state in consideration (and the respective operator dressing), they have the same correlators and thus span the same subspace. While these are technically different subspaces of the total Hilbert space, they are isomorphic. We don't expect backreaction to change any of these statements significantly.

Furthermore, note that all the operators $\Phi(X_{\Hold})$ are simple operators, in the sense that one can use the usual HKLL dictionary. However, not all the operators $\Phi(X^{t_W=0}_B)$ are in causal contact with the boundary causal domain. So, even if the subspaces are technically the same, in one case the algebra of operators which act in the subspace is made out of simple operators, while in the other there are some ``complicated'' operators. This distinction seems rather arbitrary and it is just a consequence of restricting the definition of simple operators to Heisenberg operators evolved with a particular Hamiltonian.

In order for this whole story to make sense, it is crucial that $\Phi(X_B^{t_w=0}) \not = \Phi(X_{\Hold})$. Even if these operators are background dependent but state independent (within the family of states), they are different because $X_B^{t_w=0} \sim X_{\Hold}$ implies that they will be at a different proper distance to the boundary (because one intersects a shockwave). We expect the same to be true for the FG dressed operators, even if $\Phi(X_{FG})$ is state independent. Since  $X_{FG} \sim X_{\Hold}$ are at different affine distances from the boundary,  thus $\Phi(X_{FG}) \not = \Phi(X_{\Hold})$ for $X_{FG} \sim X_{\Hold}$.

\subsection{Resummation}\label{sec:resumtime}

\begin{figure}
\begin{center}
\includegraphics[width=0.8\textwidth]{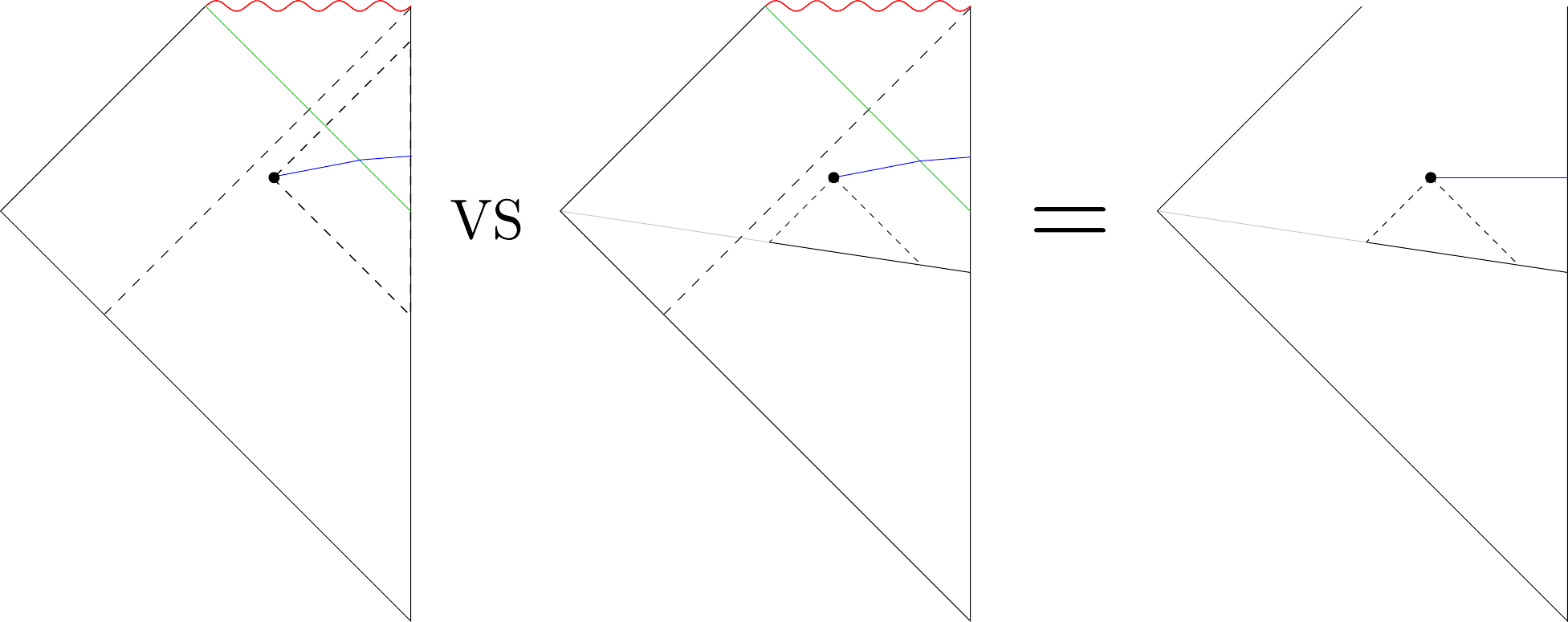}
\caption{Two methods of implementing the HKLL perscription for an operator in region I. One can either implement the usual HKLL directly by finding the spacelike Green's function in the Vaidya geometry (left), or by evolving the operator using a retarded Green's function fitting the operator entirely in region I (right). The operator can then be obtained by evolving forwards using $\Hold$. The equality on the right is equation \ref{eq:newvsoldv2}. }\label{I}
\end{center}
\end{figure}

In this subsection, we want to illustrate a confusing point. Whenever one can follow the HKLL procedure in two distinct ways for the operators, see figure \ref{I} for an illustration, it would appear that the two expressions do not match. For operators in region I in Vaidya, for example, we can map this operator to the boundary using HMPS \cite{HMPS} or using HKLL. The two distinct expressions are:
\begin{eqnarray}
e^{i \Hold t} \Phi^{\rm HMPS}(X^t_B) e^{-i \Hold t}&=& W \left [ \int d^{d-1} x' d\told' K_{\Hold}(X_{\Hold},\told(t)-\told'|x') O(x',\told')+\mathcal{O}(G_N) \right ] W^{\dagger} \nonumber \\
e^{i \Hold t} \Phi^{\rm HKLL}(X^t_B) e^{-i \Hold t}&=&\int d^{d-1} x' dt' K_H(X^t_B,t|x',t') O(x',t')+\mathcal{O}(G_N) \label{HKLLvsHMPS}
\end{eqnarray}
where we used the fact that, for Vaidya, $U(t,0)=e^{-i \Hold t} W$. Note that $\Phi(X)$ are Schrodinger operators and thus the operators in the left hand side are in the interaction picture. 
One of the goals of \cite{HMPS} was to argue that these two expressions are equivalent. But it would naively appear that these two expressions can not be equivalent, given that the right hand sides of each expression depend on different Hamiltonians and, furthermore, the operators are dressed differently.  The resolution is that the unitaries in the expression for $\Phi^{\rm HMPS}(X^t_B)$ imply that equality does not hold if we truncate to some finite order in $G_N$ . This means that, in principle, one has to resum the $\mathcal{O}(G_N)$ corrections in the first term in order to get the explicit expression of the second line. In the next section, we will consider this explicitly in $d=1$ and $d=2$ where this resummation amounts to a boundary diffeomorphism. This is of course expected from the usual way that we think about resummation. Since the operator between brackets to leading order in the first expression does not know about the shockwave, by conjugating it by the unitary the stress tensors that appear inside the bracket order by order in the $G_N$ expansion must pick up an expectation value after $t>0$ and this series resums to the HKLL answer.

Note that the equality between the two expressions can be exploited in different domains. The HKLL expression is useful for computing correlators of bulk operators in the exterior of the horizon, either before or after the shockwave, whereas the HMPS expression is useful for computing correlators of bulk operators when the endpoint $X_B^t$ is before the shockwave:
\begin{equation}
\langle \Phi^{HMPS}(X_B^t) ... \Phi^{HMPS}(Y_B^t) \rangle_{t}=\langle \Phi(X_{\Hold}) ... \Phi(Y_{\Hold}) \rangle_{\told}
\end{equation}
where the operators could in principle be inside the horizon and their geodesic could cross the shockwave, as long as the $X_i$ are in the past of the shockwave. This follows trivially from the precursor formula in \eqref{HKLLvsHMPS}, which is true in the all orders in $G_N$ sense.

\subsection{Modular flow}

Up until now we have described a procedure to reconstruct operators in the entanglement wedge by exploiting the ideas of \cite{HMPS}. However, it has recently been argued \cite{Jafferis:2015del,Faulkner:2017vdd} that one can reconstruct operators in the entanglement wedge  in terms of \emph{modular flow}. In this section we are going to explore modular flow in the shockwave states we have considered so far and make connections with the approach presented in previous sections.

The modular Hamiltonian of a given state $\rho$ is (minus) its logarithm, and \emph{modular flow} is the operation of conjugating some operator by an exponential of the modular Hamiltonian:
\begin{equation}
\mathcal{K}_{\rho}\equiv-\log \rho,\quad\quad O(x) \rightarrow O_{\sold}(x)\equiv e^{i \mathcal{K}_{\rho} \sold} O(x) e^{-i \mathcal{K}_{\rho} \sold}~.
\end{equation}

For shockwaves inserted in the thermal state $\rho_U=U \rho_{\beta} U^{\dagger}$, the modular Hamiltonian is just the thermal Hamiltonian conjugated by a unitary:\footnote{To see that this is true, note that  $U \rho^n U^{\dagger}=\left(U \rho U^{\dagger}\right)^n$~.}
\begin{equation}
\mathcal{K}_{\rho_U}=\beta U \Hold U^{\dagger}~.
\end{equation}

 We are going to consider modular flow in the Schr\"{o}dinger picture. Note that, since the modular Hamiltonian is built out of the state,  both the state and modular Hamiltonian are time dependent in this picture. We will label this time dependence with a subscript $t$ where appropriate.

In \cite{Jafferis:2015del,Faulkner:2017vdd}, the equivalence between bulk and boundary modular flows was used to write an expression for the bulk field in terms of boundary modular evolved operators:\footnote{Those papers usually refer to the $\sold=0$ version of this formula, but because modular evolution is time independent, it is easy to see that this version is also true.}
\begin{equation}
\Phi(X,\sold)=\int d^{d-1} x \int d\sold' K_{t}^{\sold-\sold'}(X|x) \rho_t^{-i \sold'} O(x) \rho_t^{i \sold'}+\mathcal{O}(G_N)~. \label{modhkll}
\end{equation}
We refer readers looking for a discussion on the intuition behind \eqref{modhkll} to \cite{Jafferis:2015del,Faulkner:2017vdd}. Since we are considering the whole state of the CFT (and not some reduced density matrix), the field $O(x)$ is therefore integrated over a boundary Cauchy slice $\partial \Sigma_t$. This formula should be valid for bulk points $X$ in $\Sigma_t$ anywhere in the entanglement wedge. Note that \eqref{modhkll} has us integrate over $\sold$, but not over $t$. 

A very simple illustration of how (\ref{modhkll}) acts can be given for the operator at the RT surface (bifurcartion horizon).  As was shown in \cite{Faulkner:2017vdd}, we expect:
\begin{equation}\label{zeromode}
\Phi(X_{RT})= \int d^{d-1} k e^{-i k x} \int d\sold\, c^{-1}_k \rho_t^{-i \sold} O_k \rho_t^{i \sold}~,\quad\quad\quad O_k=\int d^{d-1} x\,O(x) e^{i k x}~,
\end{equation}
where $O_k$ is the spatial fourier mode and $c_k$ is the fourier transform of the bulk-to-boundary correlator: $\langle \Phi(X_{RT}) O(x) \rangle$. Equation \nref{zeromode} tells us that the bulk operator at the horizon, for an arbitrary geometry, is given by a linear combination of modular ``zero modes.''  Generically, the kernel of \nref{modhkll} will be quite complicated as it depends on $\sold$ and on $\mathcal{K}_\rho$. However for $\Phi(X_{RT})$  it only depends on the bulk-to-boundary correlator.

\subsubsection{Vaidya}

Let us now consider a concrete simple example to get an easy visualization of the flow: a case similar to Vaidya, where the state has been time evolved with a Hamiltonian which is time dependent after some time $t=0$, and introduces shockwaves. As a reminder, $U=U_H(t,0)$ and we will denote evolution by the old Hamiltonian by \ $U_{\Hold}(\told,0)$.
In this setting, it is easy to understand how modular evolution acts on the bulk operator $\Phi(X^t_B)$. The modular flow will be:\footnote{For simplicity, we are going to rescale $\sold \rightarrow \sold/\beta$, so that we can identify $\sold$ in the original thermal state with the standard time evolution.}
\begin{equation}\label{modvaidya}
\Phi(X^t_B,\sold)=U_H(t,0) e^{i \Hold \sold} U_H(t,0)^{\dagger} \Phi(X^t_B) U_H(t,0) e^{-i \Hold  \sold} U_H(t,0)^{\dagger}~.
\end{equation}

\bigskip{\noindent\bf Modular flow of bulk operators}

\begin{figure}
\begin{center}
\includegraphics[width=\textwidth]{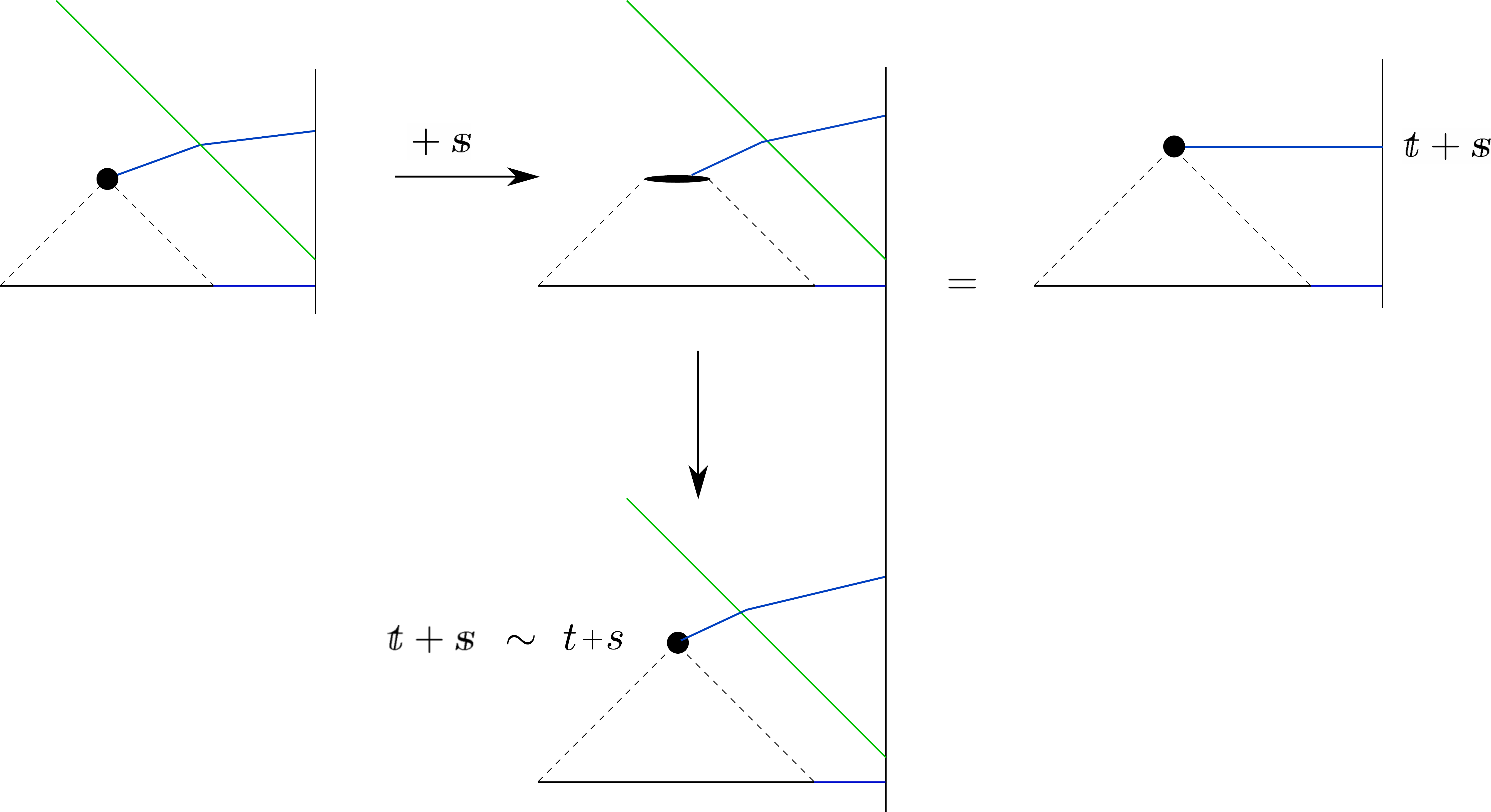}
\caption{By modular evolving the operators $\Phi(X_B^t)$ by an amount $\sold$, we get a non-local operator in the same Cauchy slice. If we Heiseneberg evolve this operator, it is equivalent to the Heisenberg operator $\Phi(X_{\Hold},\told+\sold)$ (right). If the point $(X_B',t+s) \sim (X_{\Hold},\told+\sold)$ is before the shockwave, this implies that the modular flow is local. That is, $\Phi(X_B^t,\sold)$ is equivalent to the Heisenberg operator $\Phi({X'}_B^{t+s},t+s)$ (bottom).}\label{Y}
\end{center}
\end{figure}

If we stick to the region before the geometry changes (e.g. regions I and II in Vaidya), we can use equation (\ref{eq:newvsoldv2}) to write \eqref{modvaidya} as:
\begin{equation}
U_H(t,0)^{\dagger} \Phi(X^t_B,\sold) U_H(t,0)  =U_{\Hold}(\told+\sold,0)^{\dagger} \Phi(X_{\Hold}) U_{{\Hold}}(\told+\sold,0)~,\quad\quad\quad X_B^t \sim X_{\Hold}~.   \label{modflow}
\end{equation}
The term inside (\ref{modflow}) is the Heisenberg evolved operator $\Phi(X_{\Hold},\told+\sold)$. In this way, even if the modular flow in the new state is non-local, its action on operators whose endpoint is in the past of the shockwave will be local in the old geometry. That is, if we do modular flow $\sold$ of the Heisenberg operator at time $t$ and we write the corresponding operator at $t=0$, we get the Heisenberg evolved operator with respect to $\Hold$ evolved for a time $\told+\sold$, as shown in figure \ref{Y}.

 Now, if the endpoint of the geodesic $(X_B^{\told+\sold},\told+\sold)$ is also before the shockwave,  we can again use (\ref{eq:newvsoldv2}) and (\ref{modflow}) to write :

\begin{equation}
U_H(t,0)^{\dagger} \Phi(X_B^{t},\sold) U_H(t,0)  =U_H(t+s,0)^{\dagger} \Phi(X_B^{t+s}) U_H(t+s,0)~,\quad\quad X_B^{t+s} \sim X_B^{\told+\sold}~, \label{modflowloc}
\end{equation}
where $X_{B}^{t+s}$ sits at the same proper distance along the same geodesic thrown from the RT surface as $X_B^{\told+\sold}$. That is, when the points $(X_B^t,t)$ and $(X_B^{t+s},t+s)$ are both in the past of the shockwave, we have shown the modular flow of the bulk operator $\Phi(X^t_B)$ (whose dressing crosses the shockwave) is local! See figure \ref{Y}. This justifies a posteriori the choice of this particular dressing, because it is only for this dressing that the modular flow presents these nice properties: in these coordinates, it is just a shift in the time label.\footnote{For operators dressed with FG geodesics, we expect the mapping between operators in the old and new geometry to make the relationship much more complicated.} If the point $(X_B^t,t)$ is before the shockwave but $(X^{t+s}_B,t+s)$ is not, then, the modular evolved operator won't be local but rather a  precursor (\ref{modflow}).

When the bulk operators are located after the shockwave but in the exterior of the horizon (e.g. region III in Vaidya), we expect the modular flow to also be local. This is because we expect that the action of the modular Hamiltonian on these operators to be approximately thermal at a larger temperature. Further understanding this seems complicated, but we expect that one can use the results of \cite{Anous:2016kss} to see this explicitly in AdS$_3$ (at least in perturbation theory).

Modular flow in region IV will be more complicated, but we expect that we can think of operators located in this region as a linear combination of bulk operators before the shockwave or outside the black hole. Since we understand how modular flow acts on these latter operators, this should give a consistency condition on the bulk modular flow for operators in region IV.

\bigskip
{\noindent\bf Reconstruction from  (\ref{modhkll})}

We have tried to stress that resummation is important when considering the morally equivalient but different ways one can express a bulk operator (for example a bulk operator in a Cauchy slice $\Sigma_t$ which crosses the shockwave in the exterior of the horizon). Given this consideration, it is clear that an explicit evaluation of  (\ref{modhkll}) will turn out to be complicated. 

For operators in region III, we expect that the action of the modular flow in (\ref{modhkll}) to be roughly local and thus can be  understood in terms of the usual HKLL dictionary.

For operators before the shockwave, we can  plug the explicit expression of the modular Hamiltonian in  (\ref{modhkll}). This expression includes $U_{t}=U_H(t,0)$ which is suggestive of a precursor/ resummed approach.
If we plug the modular Hamiltonian in (\ref{modhkll}) for points $X^t_B$ in regions I or II we get:

\begin{align}
U_t^{\dagger} \Phi(X^t_B,\sold) U_t& =& \int d^{d-1} x \int d\sold'  e^{i \Hold \sold'} U_{t}^{\dagger} \left[ K^{\sold-\sold'}_t(X_{\Hold}|x) O(x)+ \mathcal{O}(G_N) \right]    U_{t} e^{-i \Hold \sold'}
\end{align}
In order to evaluate this expression, which is similar to (\ref{HKLLvsHMPS}), we have to resum. Given the previous discssions, we expect
\begin{align}
U_t^{\dagger} \Phi(X^t_B,\sold) U_t=\int d^{d-1} x \int d\sold'   \tilde{K}^{\sold-\sold'}_t(X_{\Hold}|x) e^{i \Hold \sold'}  O^{\rm resum}(x,t) e^{-i \Hold \sold'}+\mathcal{O}(G_N)
\end{align}

where, again, $X^t_B$ is either in regions I or II and $X^t_B \sim X_{\Hold}$.
Consistency with equation (\ref{modflow})  requires:
\begin{equation}
   \tilde{K}^{\sold-\sold'}_t(X_{\Hold}|x) e^{i \Hold \sold'}  O^{\rm resum}(x,t) e^{-i \Hold \sold'}=  K_{\Hold}(X_{\Hold},\sold-\sold'|x) O(x,\sold'-\told(t))
\end{equation}
Where we have used the old HKLL expression for the RHS of (\ref{modflow}) and have shifted the $t$ dependence of the kernel to the operator. This implies that the operator $O^{\rm resum}(x,t)$ is the Heisenberg operator evolved with the old Hamiltonian:
\begin{equation}
O^{\rm resum}(x,t) \equiv e^{-i \Hold \told(t)} O(x) e^{i \Hold \told(t)} , \label{resumid}
\end{equation}
and the kernels are the same
\begin{equation}
\tilde{K}^{\sold-\sold'}_t(X^t_B| x)= K_{\Hold}(X,\told(t)+\sold-\sold'|x')
\end{equation}

These expressions might seem odd, but are no different than those in equation (\ref{HKLLvsHMPS}). In section \ref{sec:explicit} we will discuss how this resummation works when we can do it explicitly. In this way, the modular flow has naturally incorporated the story of the precursors and resummation. Note that generally, we usually want to keep the formula (\ref{modhkll}) as it is since the modular evolved operator $O(x,\sold)$ will be a complicated operator that does not have a particularly nice description.  However, in our particular case, because of (\ref{HKLLvsHMPS}), the resummed modular HKLL expression  is simple when evolved back to $t=0$. Note that this expression makes clear that the modular Hamiltonian acts locally on points in the old geometry. However, even in Vaidya, it is not clear at the moment how one can use modular flow to write an explicit expression for the bulk operator after the shockwave.

\subsubsection{More general geometries}

More generally, we expect the modular flow to be complicated, yet have enough structure to be able to reconstruct any operator in the entanglement wedge.  In other words, we expect any of the complicated operators that appear when evolving back and forth to be encoded in $O(x,\sold)$. We must stress, however, that whenever we want to compare the modular flow expression with an expression which has boundary unitaries, we will need resummation. In the cases we have considered, i.e. states built by acting with shockwaves on the thermal state, the modular Hamiltonian is the original Hamiltonian $\Hold$ conjugated by unitaries which act geometrically on the state. This makes the analysis simpler (as we have illustrated for Vaidya). Generically, the modular Hamiltonian is completely non-local, and we don't expect the analysis to be so simple.

From the modular flow point of view, it is not necessary to interpret the unitaries appearing in $U \rho_{\beta} U^{\dagger}$ as time evolution. For example, if we try to reconstruct operators in a time slice $\Sigma_{t>0}$ in reflecting Vaidya, we expect the expression in terms of modular flow to be exactly identical to that in Vaidya (since the modular Hamiltonian and the time slice are the same) without the need to talk about changing the boundary Hamiltonian.

In this way, modular flow provides a more natural, yet still complicated, way of thinking about precursors.

\section{Explicit examples}
\label{sec:explicit}

In this section we will implement the ideas of the previous sections in the simplified setting of two and three bulk dimensions. We will show the equivalence of the pull-back/push-forwards strategy for reconstruction in time dependent states, and furthermore show the need for resummation due the macroscopic change in the bulk geometry.

\subsection{Bulk reconstruction in AdS$_2$}\label{sec:explicitads2}

This discussion can be made most explicit in 1+1 bulk dimensions. In order to have non-trivial bulk states and dynamics, we will consider the dilaton-gravity model of JT \cite{Jackiw:1984je,Teitelboim:1983ux} and which has attracted a lot of attention recently \cite{Almheiri:2014cka,Engelsoy:2016xyb,Maldacena:2016upp,Jensen:2016pah} . The action of this model is
\begin{align}
S = {1 \over 16 \pi G_2} \int d^2 x \sqrt{-g} \left( D^2 R + C (D^2 - D^2_0)   \right)
\end{align}
where $D^2$ is the dilaton and $g$ is the two dimensional metric. The constants $C$ and $D^2_0$ parameterize the space of the theories. This theory has no bulk propagating degrees of freedom; there are four degrees of freedom all of which can be removed by two diffeomorphisms and two constraints. There is a dynamical boundary degree of freedom \cite{Engelsoy:2016xyb,Maldacena:2016upp,Almheiri:2014cka,Jensen:2016pah}. As described by the previous references, this  boundary degree of freedom can be thought of as the trajectory of the bulk cut-off surface within an unperturbed AdS$_2$ spacetime. This theory arises by looking at the near horizon limit of near extremal black holes in any dimension and truncating to the s-wave sector \cite{Almheiri:2016fws}.

We can add a bulk propagating degree of freedom by adding a scalar field term to the action. The system now describes the interaction between the scalar matter stress energy and the boundary propagating degree of freedom. For simplicity, we consider the free scalar action
\begin{align}
S_{\rm scalar} = { 1\over 16 \pi G_2} \int d^2 x \sqrt{-g} \left( {\Omega(D^2) \over 2}  \left( \nabla \Phi  \right)^2\right)
\end{align}
where $\Omega(D^2)$ is determined by the specific uplift of this model to higher dimensions. Again, for simplicity we will consider the case of $\Omega(D^2) = 1$, corresponding to coupling the scalar field only to two dimensional gravity. Since there is no direct coupling between the dilaton and the scalar field, their interaction comes completely from the gravitational constraints.

\begin{figure}
\begin{center}
\includegraphics[width=0.5\textwidth]{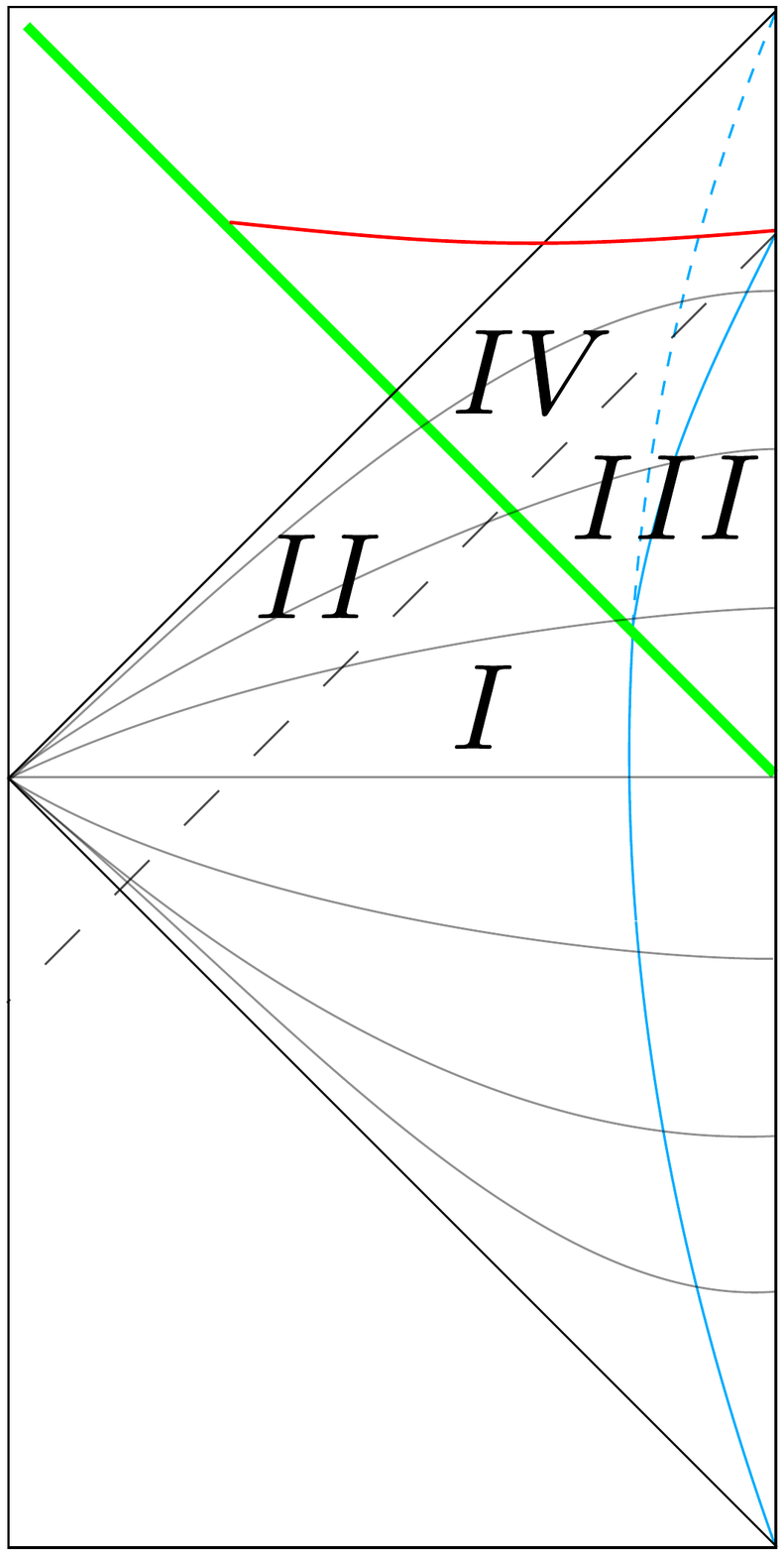}
\caption{AdS$_2$ Vaidya geoemtry. The shockwave (green) is inserted from the boundary and falls to form a black hole horizon (dashed) and singularity (red). The original AdS$_2$ Poincare slicing (grey) covers the entire spacetime. This same diagram applies of a shockwave falling into a black hole  where now the Poincare horizon can be thought of as its horizon. The Roman numerals label the regions defined earlier.}\label{shockwave}
\end{center}
\end{figure}

We want to implement the HKLL construction on the analog of the Vaidya geometry in Poincare AdS$_2$. We will work entirely in conformal gauge. This solution can be obtained by turning on the stress energy profile $T_{\tilde{v}\tilde{v}} = {\mu \over 8 \pi G_2} \delta(\tilde{v})$ and $T_{\tilde{u}\tilde{u}} = 0$ corresponding to an infalling ``shell" of matter, eventually forming a black hole. The gravity solution is given by
\begin{align}
ds^2 &= - 4 {d\tilde{u}d\tilde{v} \over (\tilde{u} - \tilde{v})^2} \label{kruskal}\\
D^2 &= D_0^2 + {a - \mu\, \Theta(\tilde{v}) \tilde{u}\tilde{v} \over \tilde{u}-\tilde{v}} \label{kruskaldil}
\end{align}
Where $a$ parametrizes a family of solutions and $\tilde{u}\equiv \tilde{t}+\tilde{z}$ and $\tilde{v}\equiv\tilde{t}-\tilde{z}$. In order for the spacetime to have a spacelike singularity we require that $\mu > 1/a$, which we will assume. The singularity occurs when $D^2 = 0$. These coordinates cover the entire Poincare patch, including the region behind the black hole event horizon; we will refer to these as Kruskal coordinates. This background is shown in figure \ref{shockwave}.

The dynamics of the gravitational sector of this model is governed entirely by the evolution of the boundary mode, or the trajectory of the cut-off surface. As shown in figure \ref{shockwave} the cut-off trajectory (blue) is perturbed by the insertion of stress energy via the shockwave (green) which causes it to terminate prematurely on the boundary. As discussed in \cite{Engelsoy:2016xyb,Maldacena:2016upp,Almheiri:2014cka,Jensen:2016pah}, this evolution of the cut-off surface tracks how the bulk and boundary times are related.

 The Kruskal coordinates \nref{kruskal} are the two dimensional analogue to the uniformizing coordinates of \ref{sec:banadosresum}. It will actually be simpler here to consider FG coordinates, unlike other sections:\footnote{Here we are not going to be too careful about the \ext gauge. The reader should keep in mind that this bulk time is not the time in \ext gauge.}
\begin{align}
&ds^2=\frac{1}{z^2} (dz^2-f(z,t) dt^2), \quad\quad\quad\quad f(z)=\Theta(-\tilde{v})+ \Theta(\tilde{v})\left(1- \frac{\mu}{a}z^2\right)^2 \nonumber \\
& \ D^2=D_0^2+\frac{a+\Theta(\tilde{v}) \mu z^2}{2 z}
\end{align}
which are related to the previous \nref{kruskal} coordinates by:
\begin{align}
  z= &\frac{1}{2}(\tilde{u}-\tilde{v}), &  t= & \frac{1}{2}(\tilde{u}+\tilde{v})~, &\tilde{v}<0 \nonumber \\ z=& \sqrt{\frac{a}{\mu}} \tanh \left[\frac{1}{2}\sqrt{\frac{\mu}{a}} \left(g^{-1}[\tilde{u}]-g^{-1}[\tilde{v}]\right)\right],  & t= & \frac{1}{2}\left(g^{-1}[\tilde{u}]+g^{-1}[\tilde{v}]\right)~, & \tilde{v}>0
\label{changeofcoord}
\end{align}
where we introduced the function $g[y]$
\begin{equation}
 g[y]\equiv \sqrt{\frac{a}{\mu}}\tanh\left( \sqrt{\frac{\mu}{a}} y\right)~,\quad\quad\quad g^{-1}[y]\equiv\sqrt{\frac{a}{\mu}}\tanh^{-1} \left(\sqrt{\frac{\mu}{a}} y\right) \label{gfunction}
\end{equation}

 These are FG coordinates where the physical boundary sits at $z=\epsilon$, which we will henceforth call the $\epsilon$-surface. Note that this surface is different from the $\tilde{z}=\epsilon$-surface ($\tilde{\epsilon}$-surface) in the Kruskal coordinates $\lbrace \tilde{u},\tilde{v} \rbrace =\tilde{t} \pm \tilde{z}$ of \nref{kruskal}. The metric in the $\tilde{z},\tilde{t}$ coordinates is AdS$_2$ in Poincar\'{e} coordinates:
\begin{equation}
ds^2=\frac{d\tilde{z}^2-d\tilde{t}^2}{\tilde{z}^2} \label{eq:unif}~
\end{equation}
and is the 2d analogue of \eqref{eq:bangroundstate} of section \ref{sec:banadosresum}. Note that these $\{\tilde{z},\tilde{t}\}$ coordinates cover regions inside the black hole and beyond the singularity.

The physically relevant $\epsilon$-surface is a complicated function of $\{\tilde{z},\tilde{t}\}$ and coincides with the $\tilde{\epsilon}$-surface at all times only when $\mu=0$.  The cut-off $\epsilon$ (solid blue) and $\tilde{\epsilon}$ (dashed blue) surfaces are depicted in figure \ref{shockwave}. Although they differ at late times, the $\tilde{\epsilon}$ and $\epsilon$-surfaces coincide before the appearance of the shockwave. In AdS$_2$, the boundary coordinate is one dimensional and is described by a timelike worldline. Therefore the different surfaces are given by different time parametrizations. For example, the $\tilde{\epsilon}$-surface gives us the physical boundary had we not perturbed the geometry with the shockwave, and the respective boundary time along this surface is the unperturbed $\told$. In keeping with our nomenclature, we denote the time along the physical $\epsilon$ boundary as $t$. In this way, evolution using $\Hold$ will give us a bulk operator in terms of operators smeared on the $\tilde{\epsilon}$ surface instead of the physical $\epsilon$ surface .

Now let us implement the HKLL prescription in various regions of the bulk. This was done for regions I and III in AdS$_2$ in \cite{Lowe:2008ra}. We will illustrate how one can also reconstruct beyond the causal horizon using the ideas developed in the previous sections. For convenience, we will consider a massless scalar field.  The scalar wave equation is
\begin{align}
\partial_{\tilde{u}} \partial_{\tilde{v}} \Phi = 0, \label{2dwaveequation}
\end{align}
for any coordinate system in conformal gauge. We will focus first on the \nref{eq:unif} coordinates which cover the entire the spacetime. In conformal gauge, we can compute the spacelike Green's function from \nref{2dwaveequation} \cite{Lowe:2008ra}:
\begin{align}
G_S = {1 \over 2}  \Theta(\tilde{z} - \tilde{z}') \Theta(\tilde{z} - \tilde{z}'-|\tilde{t} - \tilde{t}'|),
\end{align}
and the smearing function is simple to compute by projecting it onto the various surfaces of interest. For the $\tilde{\epsilon}$-surface, it is given concisely by:
\begin{align}
K(\tilde{t},\tilde{z}; \tilde{t}') = {1 \over 2} \Theta(\tilde{z} - |\tilde{t} - \tilde{t}'|)~.
\end{align}

Therefore, \ if we project this smearing function onto the $\tilde{\epsilon}$ surface, the bulk operator is given by
\begin{align}
\Phi(\tilde{t},\tilde{z}) =\int d\tilde{t}' K(\tilde{t},\tilde{z};\tilde{t}') O(\tilde{t}')= {1 \over 2} \int_{\tilde{t} - \tilde{z}}^{\tilde{t} + \tilde{z}} d\tilde{t}' O(\tilde{t}') \label{HKLLKruskal}
\end{align}
Note that this form is the same for all coordinate systems in conformal gauge, as long as we keep the $\tilde{\epsilon}$ surface fixed.As previously discussed, the meaning of the time arguments in the above expression depends on the choice of  boundary. The physically relevant boundary is the $\epsilon$ surface, but we will keep track of the $\tilde{\epsilon}$ surface to illustrate various points. For bulk operators sufficiently early in region I, the
$\tilde{\epsilon}$ and $\epsilon$ surfaces coincide and we can use the expression \nref{HKLLKruskal}. That is, the time argument of the operator $O(\tilde{t}')$ denotes the physical boundary time $t=\tilde{t}=\told$.

For operators in region III, the two surfaces are different but since the kernel is a theta function, we can keep track of the $\epsilon$-surface by simply integrating the boundary operator over the corresponding interval:
\begin{align}
\Phi(t,z) &= \int dt' K(t,z;t') O(t') = {1 \over 2} \int^{g[\tilde{u}(z,t)]}_{g[\tilde{v}(z,t)]} dt'  O(t') \label{eq:reg3}
\end{align}
where $g[\tilde{u}(z,t)]$, given in \nref{gfunction},  is the point where the future light ray (respectively past for $\tilde{v}$) emanating from $t,z$ hits the $\epsilon$ surface. By $\tilde{u}(z,t)$ (resp. $\tilde{v}(z,t)$) we mean the expression for $\tilde{u}$ ( $\tilde{v}$) in terms  $(z,t)$ obtained by inverting \nref{changeofcoord}.

We have phrased reconstruction in this scenario as a smearing of boundary operators where one has to keep track of the appropriate boundary surfaces. Alternatively, one can equivalently describe the difference between the $\epsilon$ and $\tilde{\epsilon}$-surfaces in terms of the boundary conformal transformation:
\begin{align}
 \tilde{t} \rightarrow t=f(\tilde{t})=\Theta(-\tilde{t}) \tilde{t}+\Theta(\tilde{t}) \sqrt{a \over \mu} \tanh\left(\sqrt{\mu \over a} \tilde{t} \right)  \label{bulkbdytime}
\end{align}
From this point of view, as discussed in section \ref{sec:explicit}, \nref{eq:reg3} can be thought as a conformal transformation
\begin{align}
\Phi(t,z) &=\int d\tilde{t}' K(\tilde{t},\tilde{z};\tilde{z}') O(\tilde{t}')=\int \frac{dt'}{f'(\tilde{t})
} K\left(\tilde{t}(t,z),\tilde{z}(t,z);t\right) f'(\tilde{t}') O(t')\\
&= {1 \over 2} \int^{g[\tilde{u}(z,t)]}_{g[\tilde{v}(z,t)]} dt'  O(t')~.
\end{align}
where  we used that $O(t)$ transforms as a primary operator of weight 1 (and the kernel transforms as an operator of weight $1-h=0$ as explained before). Note that this is the same result we would have obtained starting with the wave equation in the black hole coordinates covering region III.  The interpretation of the respective surfaces as different conformal frames can be checked explicitly by comparing the extrapolate dictionaries for the two surfaces as in \nref{extrapolatebanados}.

The story continues to be roughly the same for the rest of the operators in region I whose spacelike smearing extends to the cut-off surface of region III (note that \nref{HKLLKruskal} only applied for operators in the past of $z=\epsilon,t=0$). As one might expect, the result one gets is
\begin{align}\label{newvsoldads21}
\Phi(t,z) = {1 \over 2} \int_{t - z}^{0^-} dt' O(t') + {1 \over 2} \int_{0^{+}}^{g[\tilde{u}(z,t)]} dt' O(t')
\end{align}
 where all time arguments are the physical boundary time. The first term comes from smearing before the shockwave and, unsurprisingly, the second comes from smearing after the shockwave. We have split this expression into two terms for illustration purposes, as the first term can be understood as  arising from an integral over the $\epsilon$ or $\tilde{\epsilon}$ surface.

There is an alternative expression for this operator obtained via a pull-back/push-forward scheme of \eqref{eq:newvsoldv2}: in the current language, the operator $\Phi(X_{\Hold})$ correspond to the operator projected onto the $\tilde{\epsilon}$ surface. Equation \eqref{eq:newvsoldv2} for our state $U_H(t,0)=e^{-i \Hold t} W$ then implies that
\begin{align}
 W^{\dagger} e^{-i \Hold t}\Phi(z) e^{-i \Hold t} W &= {1 \over 2} \int_{t - z}^{\tilde{u}(t,z)} d\tilde{t}' O(\tilde{t}')
= {1 \over 2} \int_{t - z}^{0^-} dt' O(t')
   + {1 \over 2} \int_{0^+}^{\tilde{u}(t,z)} d \tilde{t}'  O(\tilde{t}') \label{newvsoldads2}
\end{align}
where the first term is the same for the $\epsilon,\tilde{\epsilon}$ surface but the second is different. As we have explained before, to understand \nref{eq:newvsoldv2} and the equivalence between \eqref{newvsoldads21} and \eqref{newvsoldads2}, we have to use the background independent expression for the operator:
\begin{align}
W\left [\int_{0^+}^{\tilde{u}(t,z)} d\tilde{t}' O(\tilde{t}') +... \right ] W^{\dagger}   =  \int_{0^+}^{g[\tilde{u}(t,z)]} dt' O(t') \label{resumads2}
\end{align}
 Similar to the discussion of \ref{sec:explicit}, we can think of this ``resummation" as implementing the $t>0$ part of the conformal  transformation  \nref{bulkbdytime}. From the bulk point of view, it is clear that \nref{eq:newvsoldv2} should be true: these two expressions correspond to using the spacelike Green's function to write the bulk operator in terms of the two different $\epsilon,\tilde{\epsilon}$ surfaces.\footnote{As discussed in \ref{sec:banadosresum}, the field close to the surface will be  $\Phi = \varepsilon^h O(t)+...$ and the difference between the $\epsilon,\tilde{\epsilon}$ surfaces accounts for the proper conformal transformation of the boundary operator. } This is no different than when one writes a field in terms of fields at different Cauchy slices in its past, the expression is independent of what spacelike slice one uses.

Implementing the usual HKLL prescription for regions II and IV and expressing things in terms of the boundary physical time is more subtle. When projected to the $\tilde{\epsilon}$ surface, their expression in terms of Kruskal times is given by \eqref{HKLLKruskal}. The novel thing about these operators is that they involve smearing of operators at times $\tilde{t}$ that do not map to the boundary physical time $t$; when projecting to the $\tilde{\epsilon}$ surface, the smearing contains operators at times $\tilde{t} > \sqrt{a\over \mu}$ where $\tilde{t} = \sqrt{a \over \mu}$ corresponds to $t = \infty$ under the conformal transformation \nref{bulkbdytime}. The operators in region II can then be understood in terms of evolution by $\Hold$, which is equivalent to projecting the operator to the $\tilde{\epsilon}$ surface
\begin{align}
 W^{\dagger} e^{i \Hold t}\Phi(z) e^{-i \Hold t} W  = {1 \over 2} \int_{t - z}^{0^-} dt' O(t')   +  {1 \over 2} \int_{0^+}^{\tilde{u}(t,z)} d\tilde{t}'  O(\tilde{t}')
\end{align}
which is of course the same as \nref{newvsoldads2}, with the difference is that there is no alternative description as in \eqref{newvsoldads21}.\footnote{Perhaps there is a way of first time evolving these operators using $\Hold$ to times $\tilde{t} < \sqrt{a \over \mu}$ and then performing the coordinate trasnformation to physical times $t$.}

And finally, for region IV, we can implement reconstruction via a bulk pull-back/push-forward. This involves first writing the bulk operator in IV as operators in regions I,  II and III using a bulk retarded Green's function.  Recall that we can think of the reconstruction in regions I, and II as projecting onto the  $\tilde{\epsilon}$ surface. And reconstruction in region III has the two equivalent representations in terms of operators in $\epsilon$/$\tilde{\epsilon}$. Then we can think of reconstruction in region IV as projecting onto the $\tilde{\epsilon}$ surface, just as in the previous cases. However, given that the $\tilde{\epsilon}$ surface is very particular to $d=1$, we are also going to go through the procedure explained in \ref{sec:recon} for reconstruction in this region.

Working in Kruskal coordinates, this Green's function is
\begin{align}
G_R(x,x') = {1 \over 2}  \Theta(\tilde{t} - \tilde{t}') \Theta(\tilde{t} - \tilde{t}' - |\tilde{z} - \tilde{z}'|),
\end{align}
and so the operator is
\begin{align}
\Phi(t,z) =  {1 \over 2} \left( \Phi(t_0,-{\tilde{v}(t,z)}+\tilde{t}_0)  + \Phi(t_0,{\tilde{u}(t,z)} - \tilde{t}_0) \right)+{1\over 2} \int_{-{\tilde{v}(t,z)}+\tilde{t}_0}^{{\tilde{u}(t,z)} - \tilde{t}_0} d\tilde{z}' \partial_{\tilde{t}_0} \Phi(\tilde{t}_0, \tilde{z}')
\end{align}
where this smearing is along a constant $t_0$ Kruskal slice in the bulk passing through regions I, II and III.
To simplify matters, we are considering an operator early enough in region IV such that the retarded smearing involves no boundary operators directly. If we choose $\tilde{t}_0$ to go through the intersection between the event horizon and the shockwave, i.e. $\tilde{t}_0=\frac{1}{2} \sqrt{\frac{a}{\mu}}$, we will only get a contribution from operators in regions II, III.

At this point, we would normally split the integral into two terms: before and after the shockwave (before and after $\tilde{z}=\tilde{t}_0$) and evaluate the time derivatives using the respective reconstructions. However, given the simplicity of the kernel, it turns out to be even easier to simply calculate the integral. Notice that the combination $\int_{-{\tilde{v}(t,z)}+\tilde{t}_0}^{\tilde{t}_0}+\int_{\tilde{t}_0}^{{\tilde{u}(t,z)}-\tilde{t}_0}$ is almost what we would expect if we had two operators at $(\tilde{u},\tilde{v})=(\tilde{u}(t,z),0) \in {\rm II}$ and $(\tilde{u},\tilde{v})=(2 \tilde{t}_0,\tilde{v}(t,z)) \in {\rm III}$, respectively, up to $\Phi[\tilde{u}=2\tilde{t}_0,\tilde{v}=0]$ terms. This means that we can write:
\begin{equation}
\Phi(t,z)=\Phi[\tilde{u}(t,z),\tilde{v}(t,z)]=\Phi_{\rm II}[\tilde{u}(t,z),0]+\Phi_{\rm III}[2 \tilde{t}_0,\tilde{v}(t,z)]-\Phi[2\tilde{t}_0,0] \label{twoop}
\end{equation}
where $\Phi[\tilde{u},\tilde{v}]$ simply denotes that we write the field in terms of the $\lbrace \tilde{u}(t,z),\tilde{v}(t,z) \rbrace$ variables.
This simplification is certainly only true in the case of a massless field, illustrating that it is chiral. Because of this, it should be clear that the $\tilde{t}_0$ dependence drops out---the right moving mode contribution in \nref{twoop} should only come from the II operator and the left moving mode will come from the III operator.\footnote{Right/left moving refers to the $\tilde{u},\tilde{v}$ Kruskal coordinates and are \emph{not} $t \pm z$.} We have chosen to not work with chiral fields to be careful about the different coordinates systems.

\subsubsection*{Resummation}

We would like to understand how one might get this formula from resumming the HKLL expansion. As we discussed before, in these simple examples, different background are just equivalent to different boundary choices. Given an operator at some affine distance from the boundary $z$ and time $t$, its affine distance to the $\tilde{\epsilon}$ boundary, $\tilde{z},\tilde{t}$ will depend on the details of the geometry \nref{changeofcoord}, i.e. $\tilde{z}[z,t;\mu],\tilde{t}[z,t;\mu]$.  In this way, we have that:
\begin{equation}
\Phi(z,t)=\Phi\left(\tilde{z}[z,t;\mu],\tilde{t}[z,t;\mu]\right)=\int d \tilde{t}' K\left(\tilde{z}[z,t;\mu],\tilde{t}[z,t;\mu];\tilde{t}'\right) O(\tilde{t}')
\end{equation}

All the dependence on $\mu$ just comes from the coordinate transformation. This expression can be expanded in $\mu$, and basically we can think of the different terms in the expansion as correcting for the fact that the $\epsilon$ and $\tilde{\epsilon}$ surfaces are different.

\subsection{Bulk reconstruction in AdS$_3$}\label{sec:explicitads3}
Let us now demonstrate how our discussion works in detail in the case $d=2$. For simplicity, we are going to focus on planar Vaidya, whose metric is most simply written in ingoing Eddington-Finkelstein coordinates:
\begin{equation}\label{eq:planarvaidya}
  ds^2=\frac{1}{z^2}\left(-F(z,v) dv^2-2dvdz+dx^2\right)~,\quad\quad F(z,v)\equiv 1-\Theta(v)\left(\frac{2\pi z}{\beta}\right)^2~.
\end{equation}
These coordinates are convenient for two reasons: firstly it is clear from \eqref{eq:planarvaidya} that the CFT coordinate $x$ does not change across the shock. Secondly the null coordinate $v$ coincides with the boundary time at $z=0$.

If we perform the following piecewise coordinate transformation
\begin{equation}
  v=\begin{cases} t-z~, &v<0\\ t-\frac{\beta}{2\pi}\tanh^{-1}\left(\frac{2\pi z}{\beta}\right)~, & v>0\end{cases}
\end{equation}
the geometry in each patch becomes
\begin{equation}
  ds^2=\begin{cases}\frac{1}{{z^2}}\left(-d{t}^2+d{z}^2+d{x}^2\right)~, &v<0\\ \frac{1}{z^2}\left(-f(z)dt^2+\frac{dz^2}{f(z)}+dx^2\right)~, &v>0\end{cases}
\end{equation}
with $f(z)=1-\left(\frac{2\pi z}{\beta}\right)^2$. Note that the times before and after the shockwave are discontinuous. The geometries across each patch are related by a coordinate transformation, similar to the case of the Ba\~{n}ados geometries:
\begin{equation}\label{eq:planartrans}
\tilde{z}=z e^{\frac{2\pi x}{\beta}}, \quad\quad\quad\quad \tilde{x} \pm \tilde{t}=e^{\frac{2 \pi}{\beta}(x\pm t)} \sqrt{\left(\frac{\beta}{2\pi}\right)^2-z^2}~.
\end{equation}
Our goal in this section is to illustrate how one can think of resummation in similar way to AdS$_2$, using the fact that the geometries before and after the shock can be related via a coordinate transformation.

 Notice however that the transformation \eqref{eq:planartrans} mixes the boundary coordinates and makes the discussion of resummation in $d=2$ more difficult than the $d=1$ case.

Near $z=0$ the coordinate transformation \eqref{eq:planartrans} coincides with that of \eqref{eq:banresum}, with $f_\pm(y)=\pm\frac{\beta}{2 \pi}\exp\left(\pm\frac{2 \pi y}{\beta} \right)$ after writing the metric in terms of to the proper FG coordinate  $z\rightarrow \frac{z}{1+\left(\frac{\pi z}{\beta}\right)^2}$.

 As we discussed in the previous section, as well as section \ref{sec:banadosresum}, we can obtain the HKLL kernel in region III of Vaidya by considering the vacuum AdS$_3$ spacelike greens function and project it onto the respective codimension-$1$ timelike surface after the shock:
\begin{equation}
 \tilde{z}=\epsilon e^{\frac{2\pi x}{\beta}}=\epsilon \frac{2\pi}{\beta}\sqrt{|\tilde{x}^2-\tilde{t}^2|}
\end{equation}

Of course, we can also reconstruct the operator to the past of $t=0$, by using the vacuum reconstruction to the surface $z=\tilde{z}=\epsilon$. However, unlike the $d=1$ case, these two surfaces can't be glued smoothly across the shock. This means that we can't think of doing vacuum HKLL with a different surface.

Even if we can't think of the whole geometry as vacuum AdS, we can still see how resummation works in this case, as it did for $d=1$. As per our discussion, resummation is needed in order to compare the expression obtained using the full HKLL kernel versus using the pullback/pushforward expression, as in equation \nref{HKLLvsHMPS}.   To see this, consider an operator on the shockwave, in the exterior of the horizon. The usual HKLL expression would give us the reconstruction in the thermal state, which we can understand as projecting the vacuum kernel onto the $\tilde{z}=\epsilon e^{2\pi x}$ surface. The pullback/pushforward mechanism gives the vacuum expression for this operator, which is equivalent to projecting into the $\tilde{z}=\epsilon$ slice. These two expressions just differ by a conformal transformation. This is the $d=2$ analogue of \nref{resumads2}.

\subsubsection{Extremal geodesics in Vaidya}
We will now give a discussion on the different types of geodesic dressing that we discussed in the main text, applied to AdS$_3$. Starting from the metric \eqref{eq:planarvaidya} we can solve the geodesic equation for $(X_{FG},t)$ or $(X_S,\told)$, which differ by where and how we impose our initial conditions.

It is straightforward to extend the analysis of \cite{Hubeny:2013dea} for the case of planar-Vaidya with a shock at $t=0$. We will label the components of our spacelike geodesics as $(V(s),Z(s),X(s))$, with $s$ the affine parameter along the geodesic.  The metric \eqref{eq:planarvaidya} has a Killing vector $(\partial_x)^\mu$  with an associated conserved quantity $L$ along the geodesic. The metric also admits an approximate Killing vector $(\partial_v)^\mu$ whose conserved quantity $-E$ is constant away from the shock but jumps at the surface $v=0$ in order to ensure that $V'(s)$ is continuous along the geodesic.

For $L=0$ the geodesic equations are:
\begin{align}
  X'(s)&=0~,\\
  V'(s)&=\frac{E\,Z-Z'}{F(Z,v)}~,\\
  Z'(s)^2&= Z^2\left(E^2Z^2+F(Z,v)\right)~.
\end{align}
The difference between the $(X_{FG},t)$ geodesics and the $(X_S,\told)$ geodesics is that the former are charaterized by $s$ increasing towards the bulk with initial condition $(Z(0)=\epsilon,~V(0)=t,~X(0)=x, ~E=0)$ while the latter are characterized by $s$ increasing towards the boundary and initial condition $(Z(0)=z,~V(0)=\told-z,~X(0)=x,~E=0)$. The jump in $E$ across the shock at $v=0$ is determined by ensuring continuity of $V'$.

We can calculate the (renormalized) proper distance to the bulk point $(z,t)$, $l(z,t)$, along the respective geodesics as well as the relation between boundary and locally Rindler time $t(\told)$. For extremal geodesics these are:
\begin{equation}
  l(z,t)=\mathbb{z}(z,t)=\log\left(\frac{z}{1-\left(\frac{\pi \told}{\beta}\right)^2}\right)~,\quad\quad\quad \told(t)=\frac{\beta}{\pi}\tanh \left(\frac{\pi t}{\beta}\right)~.
\end{equation}
The expressions for $FG$ geodesics are more involved and we present them here for reference:
\begin{align}
  l(z,t)&=\log z_{FG}(z,t)=\log\left\lbrace\frac{2}{z}\left(\frac{\beta}{2\pi}\right)^2\text{csch}^2\left(\frac{\pi t}{\beta}\right)\left(\sqrt{1+\left(\frac{\pi z}{\beta}\right)^2\sinh^2\left(\frac{2\pi t}{\beta}\right)}-1\right)\right\rbrace~,\nonumber \\ \told&=\frac{\beta}{2\pi}\text{csch}\left(\frac{2\pi t}{\beta}\right)\left(\cosh\left(\frac{2\pi t}{\beta}\right)-\sqrt{1+\left(\frac{\pi z}{\beta}\right)^2\sinh^2\left(\frac{2\pi t}{\beta}\right)}\right)~.
\end{align}

\subsection{Bulk reconstruction in general dimensions}

To obtain the HKLL kernel more generally in regions causally connected with the boundary, one simply solves the scalar wave equation in the time dependent background. For the geometries in consideration, this means finding the corresponding spacelike Green's function in all patches, i.e. the AdS/Schwarzchild kernels with different temperatures (and given the symmetries of the problem, one can just focus on the spatial zero mode). One then has to glue them appropiately along the shockwaves. The added technical complication in higher dimensions arises because there are no analytic expressions for the wave functions in AdS/Schwarzchild.

A similar technical complication arises with the geodesics, which must be solved for numerically. See \cite{Hubeny:2013dea} for more details.

\section{Discussion}
\label{sec:discussion}
We would like to conclude with some comments and open questions.

\subsubsection*{The notion of simple operators}

In the context of bulk reconstruction, one often talks about simple operators \cite{Papadodimas:2012aq,Papadodimas:2013jku,Papadodimas:2015jra}: low energy operators which coincide with the extrapolate limit of the bulk fields and describe bulk perturbation theory.

In our time dependent context, given a Hamiltonian $H(t)$ and the vacuum state $|0 \rangle$, we should think of a given bulk geometry as a series of states $|\psi(t)\rangle$ related by time evolution, or, in the Heisenberg picture, as a state $|0\rangle$ and an algebra of Heisenberg operators $U^{\dagger}(t) O_i(x) U(t)$, where $O_i(x)$ are single trace operators with $\Delta \sim \mathcal{O}(1)$. Note that if we wanted to focus on any other state, we could prepare it via (possibly Euclidean) time evolution.

In this way, given a Hamiltonian, simple operators are the Heisenberg operators evolved with respect to that Hamiltonian. If the dynamics are such that event horizons are formed, simple operators can't probe beyond the event horizon because of bulk causality.

This Hamiltonian dependent notion of simple operators seems rather arbitrary and is entirely a bulk notion: the Heisenberg operator evolved with the time independent Hamiltonian, $O(x,\told)$ are clearly simpler in the boundary than a Heisenberg operator evolved with a complicated Hamiltonian, $O(x,t)$. However, it is the latter that are present in the extrapolate dictionary in the state evolved with $H(t)$. As we have described, the operators inside the horizon are given by the $O(x,\told)$ operators, which don't appear simple in the bulk.

The idea of extending the notion of ``simple'' operators to allow for Heisenberg operators
evolved with distinct Hamiltonians was also discussed in \cite{Engelhardt:2017aux}, in order to understand the entropy of marginally trapped horizons. The context is the same as in our paper, since these generalized ``simple'' operators naturally probe beyond the causal horizon.

\subsubsection*{Modular time and gauge invariance}

As we have discussed, for the class of states built by adding shockwaves to simpler states, the locality of the modular flow seems to single out the \ext foliation, a particular set of bulk Cauchy slices  where the modular flow acts locally (close to the RT surface).\footnote{It singles out this set of Cauchy slices deep in the bulk, but when thinking of the algebra of operators in that Cauchy slice, it is natural to extend them towards the boundary using spacelike geodesics.} Because of gauge invariance, we usually don't expect a preferred foliation of spacetime in a theory of gravity. However, the combination of having bulk matter and locality of modular evolution seems to pick a foliation. This appears to be a generalization of \cite{Connes:1994hv} to a situation where the modular Hamiltonian is not local everywhere but only ``deep" in the bulk. Moreover the \ext foliation can only cover part of the spacetime.

Of course, if the modular Hamiltonian is completely non-local, we don't expect the \ext foliation to provide a natural foliation of the spacetime. However, any modular Hamiltonian will be locally Rindler close to the entangling surface. We thus leave the further exploration of if one can define a preferred foliation for the entanglement wedge, starting from this intuition, to future work. It is far from clear if this is possible, because  light rays emanating from the RT surface will generically form caustics.

\subsection*{Dressing choice}
Our choice of dressing  has the property that it is simplest in the part of the spacetime which is the closest to the RT surface. This is in contrast with FG geodesics, which, as discussed in section \ref{sec:explicit}, have worse properties. This \ext slicing is the simplest geodesic slicing that we could imagine that gives a nice foliation of (part of) the spacetime.
One can certainly consider other dressings, ones whose operators are less state dependent, but we found that this usually came at the expense of their geodesics having caustics or not reaching the RT surface.

The main shortcoming of the \ext dressing is that it doesn't cover the entire spacetime. We don't view this as a fundamental limitation, but more as an indication that the operators not covered by the \ext foliation might be more complicated. It would be interesting to understand if there exists any other choice of dressing with similar nice properties to the \ext foliation, yet covers the regions invisible to the \ext foliation. The simplest candidate would be the operators dressed to the RT surface, but it is not entirely clear to us how to give a boundary prescription for them, especially for points in the region that the \ext foliation doesn't probe.

\subsection*{Operators dressed to the horizon and state dependence}

Let us elaborate a bit on the last point. While throughout we mainly worked with the operators, $\Phi(X_B,t)$, we also discussed the operators dressed to the horizon $\Phi(X_S)$ in section \ref{sec:gauge}. We would like to better understand the properties of these operators, which would appear to commute with the boundary Hamiltonian. One way to do this is to consider conjugating  $\Phi(X_S)$ by the ``dressing-changing'' operator which measures the geodesic distance from the  boundary-to-horizon spacelike geodesic along a fixed Rindler angle $\Phi(X_V)=V_{U,t} \Phi(X_S) V_{U,t}^{\dagger}$. This operator depends on $\lambda,\told,x$ and thus creates a \emph{boundary} dressed operator at fixed distance from the RT surface. This operator is different from $\Phi(X_B^t)$: even if in a given background both might be dressed to the boundary along the same geodesic. The reason is that one creates an operator at a fixed distance from the RT surface ($\lambda$) while the other does at a fixed distance to the boundary ($\mathbb{z}$)
. Note that $V$ itself depends strongly on the family of states $\rho_{\beta}$ but within this family of states the dependence in the particular geometry is just a background effect.
 This is an entanglement wedge gravitational analogue of the boundary-to-boundary Wilson line of  \cite{Guica:2015zpf,Harlow:2015lma}. We leave a more careful analysis of $V$ for future work.

It would also be nice to elaborate on the differences between defining operators inside-out as in our \ext foliation (as in $X_S,X_B^t$) versus defining them outside-in from the boundary (as in $X_{FG}$). While operators $\Phi(X_B^t)$ depend on the family of states $\lbrace \rho_{\beta} \rbrace$, they have nice local properties. Even if this state dependence is mild, one might want to consider operators that are completely state independent, such as $\Phi(X_{FG})$. Again, these operators won't have nice local properties---in order for them to stay in the entanglement wedge, one has to limit the allowed proper distances in a state and time dependent way.  We expect this tradeoff between state dependence and locality to be generic.

\subsection*{Entanglement wedge reconstruction and modular flow}
We have discussed how the proposal of \cite{Jafferis:2015del,Faulkner:2017vdd} for reconstruction in the entanglement wedge using the boundary modular flow works in our situation. In the case of Vaidya, the precursors of \cite{HMPS} are captured in terms of modular flow. Furthermore, even if the boundary modular  Hamiltonian is non-local (it is the Hamiltonian conjugated by the shockwave operator), the action of the modular flow on operators in the ``old'' part of the geometry is local. This is a boundary argument, since it is only there that we understand what the modular Hamiltonian is. We would like have a bulk understanding of why the bulk modular Hamiltonian is approximately local before the shockwave. The confusion stems because, in principle, the modular Hamiltonian depends on the whole state, whose respective Cauchy slice crosses the shockwave.

We were able to make progress because the modular Hamiltonian, even if non-local, is simple. For more general time dependent states (not built out of shockwaves), we expect it to be more complicated and won't preserve any notion of locality.

Consider the related example of a CFT in its vacuum state and consider a spherical subregion of this state: the modular Hamiltonian will be an integral of the stress tensor, similar to our ${\cal K}_{\rho_{\beta}}=\beta \Hold$. Consider acting on this state with a unitary that factorizes into the spherical subregion and its complement, the modular Hamiltonian will be just conjugated by a unitary. We could then consider modular  flow in this example and we expect our procedure to apply straightforwardly in this case.

\subsection*{Transversable wormholes}

We focused on refining the boundary expression for operators behind the event horizon. We started by focusing on the Hilbert space, in which evolution by distinct Hamiltonians appeared naturally. This is in contrast with the way discussions on reconstruction behind the event horizon are typically framed, where both the Hamiltonian and the collapsing black hole state are thought of as fixed. In contrast, in our discussion above on reconstruction behind the horizon, the Hamiltonians which appeared don't play any role in the dynamics.

In  \cite{Gao:2016bin,Maldacena:2017axo,Almheiri2017,Kourkoulou:2017zaj}, they gave a discussion on how to deform the actual boundary Hamiltonian to bring the black hole interior into causal contact with the boundary. This effect might allow us to check our expression for the interior operators. Having a deformation of the Hamiltonian which causally connects the interior operators with the boundary implies one can write them in terms of simple boundary operators using a spacelike Green's function, i.e. one can then perform the usual HKLL prescription. Since we know ultimately that the interior operators need to be complicated, perhaps one can think of them in terms of timefolds of this new deformed Hamiltonian which makes the black hole wormhole traversable. We would like to elaborate on this point in future work.

\subsection*{The trans-Planckian problem}

In order to describe the operators behind the event horizon, we used consistency of the pull-back/push-forward method of to evolve interior operators using distinct Hamiltonians and write it in terms of data at earlier times. This inherently assumes that we can propagate the mode, or it's data at earlier times, along the spacetime and through the shell which formed the black hole. In particular, this assumes that this propagation does not affect the shell nor, vice versa, does the presence of the shell preclude forward propagation of the mode (except eikonally where the shell/operator propagates on the background created by the other). As already discussed in \cite{Almheiri:2012rt,Almheiri:2013hfa,Susskind:2012uw}, the center of mass energy of this collision becomes Planckian once the modes considered are a scrambling time deep inside the black hole. This would preclude this method of reconstruction deep in the black hole interior.

\subsection*{Negative energy}

Late Rindler modes just behind black hole horizons carry negative energy with respect to the apprioximate Killing symmetry of the system at late times. This property of the modes, that they lower the asymptotic energy of the state, was argued to lead to paradoxes in \cite{Almheiri:2013hfa,Marolf:2013dba}, precluding them from being state independent operators. Our operators depend on the family of states in consideration and whether an operator is behind the horizon will be state dependent (since the proper distance from the boundary to the horizon depends on the state). It would be interesting to check (assuming we overcome the trans-Planckian problem to write late time modes) if some particular linear combinations of our operators does indeed lower the energy.